\documentclass[10pt,journal,compsoc]{IEEEtran}

\usepackage[nocompress]{cite}
\usepackage{array}
\usepackage{mdwmath}
\usepackage{mdwtab}
\usepackage{eqparbox}
\usepackage[hyphens,spaces,obeyspaces]{url}
\usepackage{amsmath}
\usepackage{float, xcolor}
\usepackage{graphicx}
\usepackage{multirow}
\usepackage{subfig}
\usepackage{booktabs}
\usepackage{amsmath,graphicx}
\usepackage{amsxtra}
\usepackage{threeparttable}
\usepackage[breaklinks=true]{hyperref}
\usepackage{bm}

\begin{document}


\title{Emotion Intensity and its Control for\\Emotional Voice Conversion}

\author{Kun Zhou,~\IEEEmembership{Student Member,~IEEE,}
        Berrak Sisman,~\IEEEmembership{Member,~IEEE,}
        Rajib Rana,~\IEEEmembership{Member,~IEEE,}\\
        Bj{\"o}rn W.\ Schuller,~\IEEEmembership{Fellow,~IEEE,}
        and Haizhou Li,~\IEEEmembership{Fellow,~IEEE}
\IEEEcompsocitemizethanks{

\IEEEcompsocthanksitem Kun Zhou is with the Department of Electrical and Computer Engineering, National University of Singapore, Singapore.
E-mail: zhoukun@u.nus.edu \protect
\IEEEcompsocthanksitem Berrak Sisman is with Singapore University of Technology and Design, Singapore.
E-mail: berraksisman@u.nus.edu \protect
\IEEEcompsocthanksitem Rajib Rana is with University of Southern Queensland, Australia. E-mail: Rajib.Rana@usq.edu.au\protect

\IEEEcompsocthanksitem Bj{\"o}rn W. Schuller is with GLAM -- the Group on Language, Audio, and Music, Imperial College London, U.\,K., and the Chair of Embedded Intelligence for Health Care and Wellbeing, University of Augsburg, Germany. Email: bjoern.schuller@imperial.ac.uk\protect

\IEEEcompsocthanksitem Haizhou Li is with the Department of Electrical and Computer Engineering, National University of Singapore, Singapore, University of Bremen, Germany, and the Chinese University of HongKong (Shenzhen), China. Email: haizhou.li@nus.edu.sg\protect


}}



\IEEEtitleabstractindextext{%
\begin{abstract}
Emotional voice conversion (EVC) seeks to convert the emotional state of an utterance while preserving the linguistic content and speaker identity. In EVC, emotions are usually treated as discrete categories overlooking the fact that speech also conveys emotions with various intensity levels that the listener can perceive. In this paper, we aim to explicitly characterize and control the intensity of emotion. 
We propose to disentangle the speaker style from linguistic content and encode the speaker style into a style embedding in a continuous space that forms the prototype of emotion embedding. We further learn the actual emotion encoder from an emotion-labelled database and study the use of relative attributes to represent fine-grained emotion intensity. To ensure emotional intelligibility, we incorporate \textit{emotion classification loss} and \textit{emotion embedding similarity loss} into the training of the EVC network. As desired, the proposed network controls the fine-grained emotion intensity in the output speech.
Through both objective and subjective evaluations, we validate the effectiveness of the proposed network for emotional expressiveness
and emotion intensity control.  


\end{abstract}

}

\maketitle

\IEEEdisplaynontitleabstractindextext

\IEEEpeerreviewmaketitle

\IEEEraisesectionheading{\section{Introduction}\label{sec:introduction}}
\IEEEPARstart{E}{motional} Voice Conversion (EVC) is a technique that seeks to manipulate the emotional state of an utterance while keeping other vocal states unchanged \cite{zhou2021emotional}. 
It allows for the projection of the desired emotion into the synthesized voice. Emotional voice conversion poses a tremendous potential for human-computer interaction, such as enabling emotional intelligence into a dialogue system \cite{pittermann2010handling,crumpton2016survey,rosenberg2021prosodic}.

Voice conversion aims to convert the speaker-dependent vocal attributes such as the speaker identity while preserving the linguistic information \cite{sisman2020overview}. Since the speaker information is characterized by the physical structure of the vocal tract and manifested in the spectrum \cite{ramakrishnan2012speech}, spectral mapping has been the main focus of voice conversion \cite{kain1998spectral}.
However, speech also conveys emotions with various intensity levels that can be perceived by the listener \cite{wang2009exploration}. 
For example, happy can be perceived as happy or elation \cite{averill1993happiness}, while angry can be divided into a `mild' angry and the `full-blown' angry \cite{biassoni2016hot}. In particular, intensity of emotion is described as the magnitude of factor to attain the goal of the emotion \cite{brehm1999intensity}. Therefore, emotion intensity is not just the loudness of a voice, but correlates to all the acoustic cues that contribute to achieving an emotion \cite{frijda1992complexity}. Moreover, speech emotion is hierarchical and supra-segmental in nature, varying from syllables to utterances \cite{tao2006prosody,rosenberg2009detecting,levitan2011measuring,schuller2013computational}. 
Thus, it is insufficient to only focus on frame-wise spectral mapping for emotional voice conversion. Both intensity variations and prosodic dynamics need to be considered for speech emotion modelling. 

Synthesizing various intensities of an emotion is a challenging task for emotional voice conversion studies. One of the reasons is the lack of explicit intensity labels in most emotional speech datasets. Besides, emotion intensity is even more subjective and complex than just considering discrete emotion categories, which makes it challenging to model \cite{frijda1992complexity}. There are generally two types of methods in the literature for emotion intensity control. One uses auxiliary features such as a state of voiced, unvoiced, and silence (VUS) \cite{matsumoto20_interspeech}, attention weights or a saliency map \cite{schnell11improving}. Another manipulates the internal emotion representations through interpolation \cite{um2020emotional} or scaling \cite{choi2021sequence}. 
Despite these methods, emotion intensity control is still an under-explored topic in emotional voice conversion. 

Previous emotional voice conversion studies mainly focus on learning a feature mapping between different emotion types. 
Most of them, model the mappings of spectral and prosody parameters with a Gaussian mixture model (GMM)~\cite{aihara2012gmm,kawanami2003gmm}, sparse representation~\cite{aihara2014exemplar}, or hidden Markov model (HMM)~\cite{inanoglu2009data}. Recent deep learning methods such as deep neural networks (DNN) \cite{lorenzo2018investigating, luo2017emotional} and deep bi-directional long-short-term memory network (DBLSTM) \cite{ming2016deep} have advanced the state-of-the-art. New techniques using generative adversarial network (GAN)-based \cite{Zhou2020, shankar2020non, rizos2020stargan} or auto-encoder-based models \cite{zhou2020converting, zhou2021vaw, gao2019nonparallel} make it possible for non-parallel training. We note that these frameworks convert the emotion on a frame basis, so speech duration cannot be modified. Moreover, since the spectrum and prosody are not independent of each other, a separate study of them may cause a mismatch during the conversion \cite{shankar2020multi, qian2020f0}. It would be advantageous to have a model to transfer the correlated vocal factors end-to-end, producing more realistic emotions in synthetic speech.

Recently, sequence-to-sequence (Seq2Seq) models have attracted much interest in speech synthesis~\cite{kyle2017char2wav,wang2017tacotron} and voice conversion \cite{zhang2019sequence, tanaka2019atts2s,kameoka2020many,kreuk2021textless}.  With the attention mechanism, Seq2Seq frameworks jointly learn the feature mapping and alignment and automatically predict the speech duration at run-time. 
Inspired by these successful attempts, researchers introduce Seq2Seq modelling into emotional voice conversion. For example, a Seq2Seq model to jointly model pitch and duration is proposed in \cite{robinson2019sequence}. In \cite{kim2020emotional}, a multi-task learning for both emotional voice conversion and emotional text-to-speech is studied. We note two limitations of these studies: First, they learn an averaged emotional pattern during the training, while emotional expressive speech presents abundant variations of emotion intensity in real life. Second, these frameworks require enormous emotional speech data to train. \emph{But in practice, such a large emotional speech database is not widely available, which limits the scope of applications.}

In this article, we aim to address the above challenges. The main contributions of this paper are listed as follows. 
\begin{itemize}
 \item We introduce \textit{Emovox}, a Seq2Seq emotional voice conversion framework, which jointly transfers the spectrum and duration in an end-to-end way for emotional voice conversion.   
 \item 
    \textit{Emovox} automatically learns the abundant variations of intensity that are exhibited in an emotional speech dataset, without the need for any explicit intensity labels and enables effective control of the emotion intensity in the converted emotional speech at the run-time;

    \item \textit{Emovox} eliminates the need for a large amount of emotional speech data for Seq2Seq EVC training and still achieves remarkable performance under limited data conditions;
    
    \item We present a comprehensive evaluation to show the effectiveness of \textit{Emovox} for emotional expressiveness and emotion intensity control.

\end{itemize}

This paper is organized as follows: In Section \ref{sec:related work}, we motivate our study by introducing the background and related work. In Section \ref{sec: main}, we present the details of our proposed \textit{Emovox} framework and we introduce our experiments in Section~\ref{sec: exp}. In Section \ref{sec: results}, we report the experimental results and conclude in Section \ref{sec: concludes}.

\section{Background and Related Work}
\label{sec:related work}

This work is built on several previous studies spanning emotion intensity, expressive speech synthesis, and emotional voice conversion. We briefly introduce the related studies to set the stage for our research and summarize the gaps in current literature to place our novel contributions.

\subsection{Emotion Intensity in Vocal Expression}
\label{sec:emotion intensity}

The most straightforward way to characterize emotion is to categorize it into several different groups \cite{whissell1989dictionary,ekman1992argument}; however, the choice of emotion labels is mostly intuitive and inconsistent in the literature. One key reason is that emotion intensity can affect our perception of emotions \cite{juslin2001impact}. For example, happy can be perceived as happy or elation, which are similar in voice quality but different in intensity \cite{averill1993happiness}. Thus, correlating the emotion intensity to the loudness of the voice is a rather oversimplification. Emotion intensity can be observed in various acoustic cues, not only in speech energy but also in speech rate and fundamental frequency \cite{frijda1992complexity}. The differences in these cue levels could be larger between different intensities of the same emotion than between different emotions \cite{juslin2001impact}.

\begin{figure}[t]
    \centering
    \includegraphics[width=0.43\textwidth]{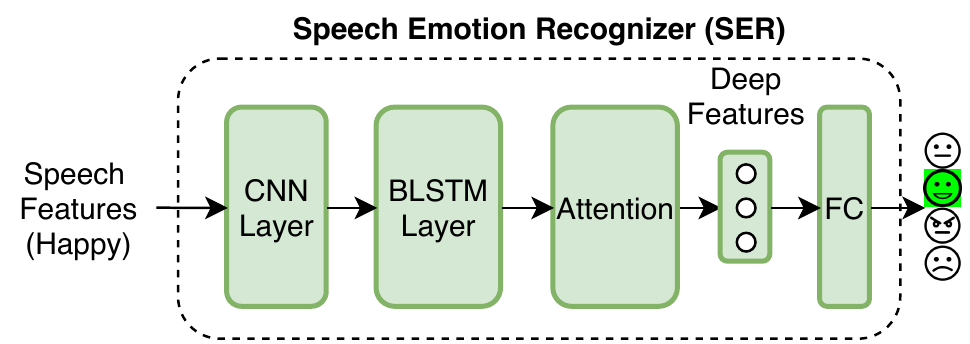}
    \caption{An example of a speech emotion recognizer (SER) \cite{chen20183}, where the deep features are obtained before the last fully-connected (FC) layer to describe the emotion styles \cite{zhou2021seen, liu2021expressive}.}
    \label{fig:ser}
\end{figure}

\subsection{Sequence-to-Sequence Conversion Models}
\label{sec:seq2seq}

The sequence-to-sequence model with attention mechanism was first studied in machine translation \cite{bahdanau2015neural} and then found effective in speech synthesis \cite{kyle2017char2wav,wang2017tacotron}. 
In text-to-speech, sequence-to-sequence modelling achieves remarkable performance by learning an attention alignment between the text and acoustic sequence, such as Tacotron \cite{wang2017tacotron}. Similar to text-to-speech, voice conversion aims to generate realistic speech from internal representations; therefore, sequence-to-sequence models are applied to various voice conversion and emotional voice conversion studies.  

\subsubsection{Sequence-to-Sequence Voice Conversion}
\label{sec: seq2seq-vc}
Sequence-to-sequence voice conversion frameworks such as SCENT \cite{zhang2019sequence}, AttS2S-VC \cite{tanaka2019atts2s}, and ConvS2S-VC \cite{kameoka2020convs2s}, jointly convert the duration and prosody components, and achieve higher naturalness and similarity than conventional frame-based methods. To address the conversion issues such as the deletion and repetition caused by the misalignment, various approaches are proposed, such as a monotonic attention mechanism \cite{kameoka2020many}, non-autoregressive training \cite{hayashi2021non,kameoka2021fasts2s}, and the use of pre-training models \cite{huang2021pretraining} or text supervision \cite{liu2021any,zhang2019non,zhang2019improving}. These successful attempts further motivate the study of sequence-to-sequence modelling for emotional voice conversion.

\subsubsection{Sequence-to-Sequence Emotional Voice Conversion}
\label{sec: seq2seq-evc}
Compared with conventional frame-based models, sequence-to-sequence models are more suitable for emotional voice conversion. First, the sequence-to-sequence models allow for the prediction of speech duration at the run-time, which is an important aspect of the speech rhythm and strongly affects the emotional prosody \cite{wu2010acoustic}. Besides, a joint transfer of spectrum and prosody in sequence-to-sequence models addresses the mismatch issues in conventional analysis-synthesis-based emotional voice conversion systems \cite{shankar2020multi, Zhou2020, gao2019nonparallel}. Also, emotional prosody is supra-segmental and can be only associated with a few words \cite{chen20183}. Learning an attention alignment makes it possible to focus on emotion-relevant regions during the conversion. Hence, sequence-to-sequence modelling for emotional voice conversion will be our primary focus in this paper.

There are only few studies on sequence-to-sequence emotional voice conversion \cite{robinson2019sequence, kim2020emotional, choi2021sequence, zhou21b_interspeech}. In \cite{robinson2019sequence}, the authors jointly model pitch and duration with parallel data, where the model is conditioned on the syllable position in the phrase. In \cite{kim2020emotional}, a multi-task learning framework of emotional voice conversion and emotional text-to-speech is built with a large-scale emotional speech database. In \cite{choi2021sequence}, the authors introduce an emotion encoder and a speaker encoder into the sequence-to-sequence training for emotional voice conversion. We note that these frameworks require tens of hours of parallel emotional speech data, which is hard to collect. A recent work \cite{zhou21b_interspeech} proposes a 2-stage training strategy for sequence-to-sequence emotional voice conversion leveraging text-to-speech to eliminate the need for a large emotional speech database. However, none of these frameworks study emotion intensity variations, and the converted emotional utterances lack the controllability of emotion intensity. Only \cite{choi2021sequence} attempts to scale the emotion embedding by multiplying it with a factor to control the emotion intensity at run-time. However, the authors do not explicitly model emotion intensity variations during the training, and their intensity control method lacks interpretability. 

This work aims to bridge this gap in the current literature and study emotion intensity modelling for emotional voice conversion. We aim to build a sequence-to-sequence emotional voice conversion framework with effective emotion intensity control using a limited amount of emotional speech data.

\subsection{Expressive Speech Synthesis with Prosody Style Control}
\label{sec: expressive}

Speech emotion is highly related to speech prosody and influenced by several prosodic cues embedded in acoustic speech such as intonation, rhythm, and energy \cite{pierre2003production,erickson2005expressive}. 
The most straightforward way to model and control prosody style is to use explicit annotations, or labels \cite{luong2017adapting,fan2015multi,henter2017principles}. Besides explicitly labelling, researchers use a reference encoder to imitate and transplant the reference style in an unsupervised way \cite{skerry2018towards}. Global style token (GST) \cite{wang2018style} is an example to learn interpretable style embeddings from the reference audio. By choosing specific tokens, the model could control the style of synthesized speech. Other studies \cite{lee2019robust, klimkov19_interspeech,li21r_interspeech,tan21_interspeech} mainly replace the global style embedding with fine-grained prosody embedding. Some other studies based on Variational Autoencoders (VAE) \cite{kingma2013auto} show the effectiveness of controlling the speech style by learning, scaling, or combining disentangled representations \cite{zhang2019learning,kenter2019chive}.

Emotion expressive speech is even more complex, which has subtle dynamic variations associated with multiple prosodic attributes \cite{murray1993toward,xu2011speech,tahon2018can}. 
Inspired by the successful attempts in prosody style control, several studies control the emotion intensity for emotional speech synthesis. For example, in \cite{um2020emotional}, an inter-to-intra distance ratio algorithm is applied to the learnt style tokens for emotional speech synthesis, where an interpolation technique is used to control emotion intensity. In \cite{schnell11improving}, the authors show that a speech emotion recognizer is capable of generating a meaningful intensity representation via attention or saliency. In \cite{zhu2019controlling,lei2021fine}, a relative attribute scheme is introduced to learn the emotion intensity for emotional speech synthesis. None of these frameworks explicitly models prosody style, but rather encodes the association between input text and its emotional prosody style end-to-end.

This contribution studies explicit modelling of emotion intensity variations with a relative attribute method for emotional voice conversion. We believe that the relative attributes scheme provides a straightforward way to model intensity variants, which will be discussed later.

\subsection{Emotional Prosody Modelling with a Speech Emotion Recognizer}
\label{sec: ser}
Emotional prosody is prominently exhibited in emotional expressive speech \cite{ong2019modeling,busso2016msp,busso2008iemocap}, which can be characterized by either categorical \cite{ekman1992argument} or dimensional representations \cite{russell1980circumplex}. Recent studies \cite{kim2019dnn} in speech emotion recognition provide valuable insights into emotional prosody modelling. Instead of categorical or dimensional attributes, they characterize the emotion styles with the latent representations learnt by the deep neural network. 
Compared with human-crafted features, deep features learnt by a speech emotion recognizer (SER) are data-driven and less dependent on human knowledge \cite{schuller2020review, latif2021survey}, which we believe is more suitable for emotion style transfer.

Some studies are leveraging a speech emotion recognizer to improve the prosody modelling for expressive speech synthesis. In \cite{gao2020interactive}, an emotion recognizer is used to extract the style embedding for style transfer. In \cite{liu2021expressive}, a speech emotion recognizer is further used as the style descriptor to evaluate the style reconstruction performance. In \cite{zhou2021seen}, researchers use the deep emotional features from a pre-trained speech emotion recognizer to transfer both seen and unseen emotion styles. These studies show the capability of a speech emotion recognizer to describe emotion styles with their latent representations.

A speech emotion recognizer also shows a potential to supervise the emotional speech synthesis system to generate the speech with desirable emotion styles \cite{schuller2018speech}. In \cite{liu21p_interspeech}, a reinforcement learning paradigm for emotional speech synthesis is proposed, where the classification accuracy of the speech emotion recognizer is used as the reward function to the system. In \cite{li2021controllable}, the authors use emotion classifiers to enhance the emotion-discrimination of the emotion embedding and the predicted Mel-spectrum. In \cite{cai2021emotion}, an emotional speech synthesis system is built on an expressive TTS corpus with the assistance of a cross-domain emotion recognizer. These studies show remarkable performance by incorporating the supervision from the pre-trained emotion recognizer into the emotional speech synthesis systems, which motivates our study. 
We further study the use of perceptual losses in EVC training to improve the intelligibility of the converted emotion.

\subsection{Research Gap (Summary)} Below, we summarise the gaps in the literature of emotional voice conversion that we aim to address in this paper.
\begin{enumerate}
    \item There are very few studies on emotion intensity control, which is crucial to achieving emotional intelligence.
    \item Despite the tremendous potential, emotion intensity control is still not a well-explored research direction for emotional voice conversion.
    \item There is a lack of focus on modelling prosody style to achieve improved emotion intensity control.
    \item Feasibility of using a pre-trained speech emotion recognizer as an emotion supervisor for EVC training poses tremendous potential but is not well understood.
\end{enumerate}

\section{\textit{Emovox}: Emotional voice conversion with emotion intensity control}
\label{sec: main}

\begin{figure}[t]
    \centering
    \includegraphics[width=0.5\textwidth]{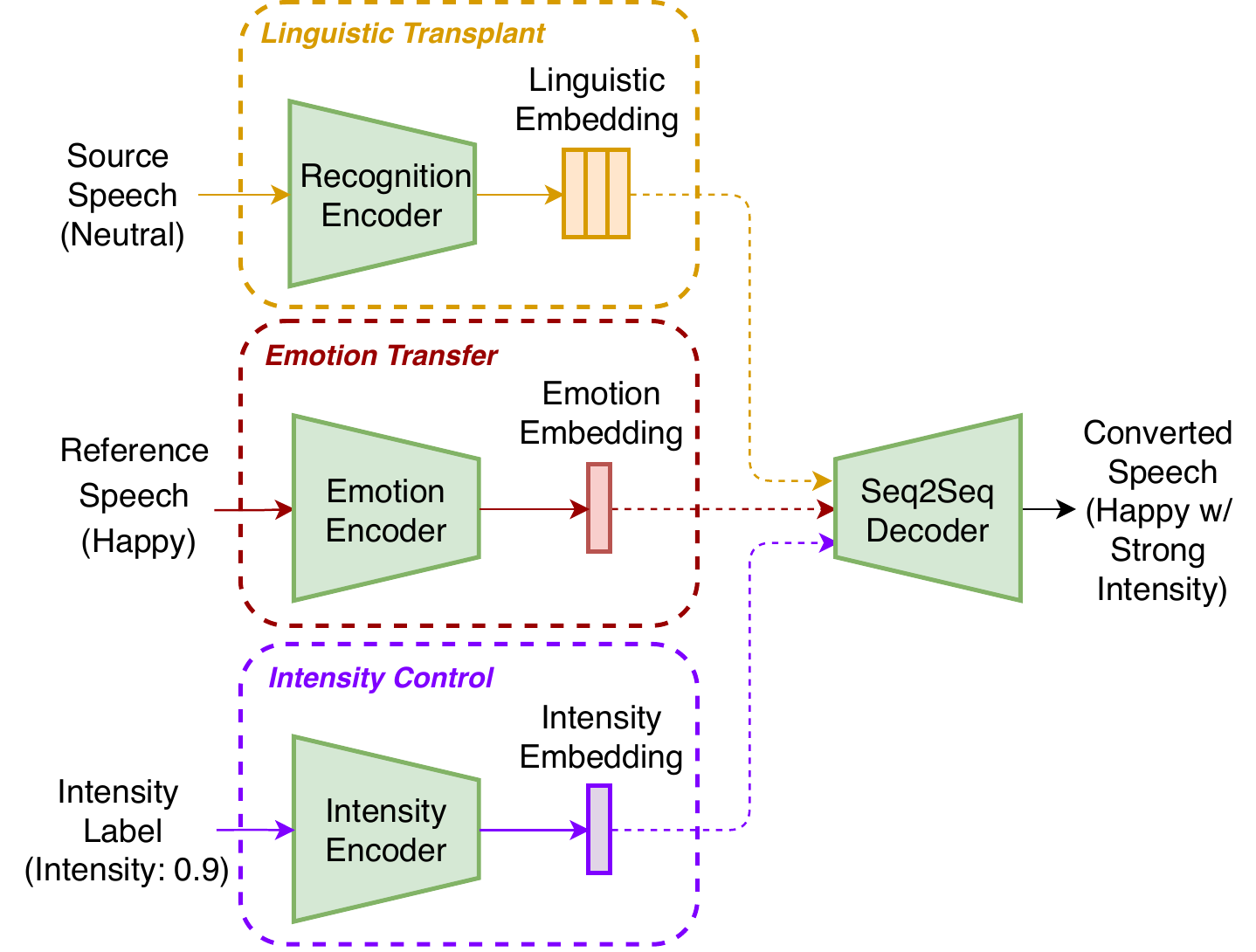}
    \caption{Block diagram of \textit{Emovox} during the conversion stage. \textit{Emovox} aims to transfer the reference emotion to the source speech (''emotion transfer'') while controlling its emotion intensity (''intensity control'') and preserving the source linguistic information (''linguistic transplant'').}
    \label{fig:overview}
\end{figure}

The proposed emotional voice conversion framework: \textit{Emovox} consists of four modules, 1) a \textit{recognition encoder}, which derives the linguistic embedding from the source speech; 2) an \textit{emotion encoder}, which encodes the reference emotion style into an emotion embedding; 3) an \textit{intensity encoder}, which encodes a fine-grained intensity input into an intensity embedding, 
and 4) a \textit{Seq2Seq decoder}, which generates the converted speech from a combination of linguistics, emotion, and intensity embeddings. At run-time,  \textit{Emovox} preserves the source linguistic content (''linguistic transplant''), while transferring the reference emotion to a source utterance (''emotion transfer''), as illustrated in Figure \ref{fig:overview}.  \textit{Emovox} also allows users to manipulate/control the emotion intensity of the output speech (''intensity control'').


To train \textit{Emovox}, we propose a Seq2Seq framework that disentangles the speech elements from input acoustic features, and reconstructs the acoustic features from the speech elements. To reduce the amount of training data for \textit{Emovox}, we introduce two pre-training strategies, i.\,e., 1) style pre-training with a large TTS corpus, and 2) emotion supervision training with an SER. 

\begin{figure*}[t]
    \centering
    \includegraphics[width=0.8\textwidth]{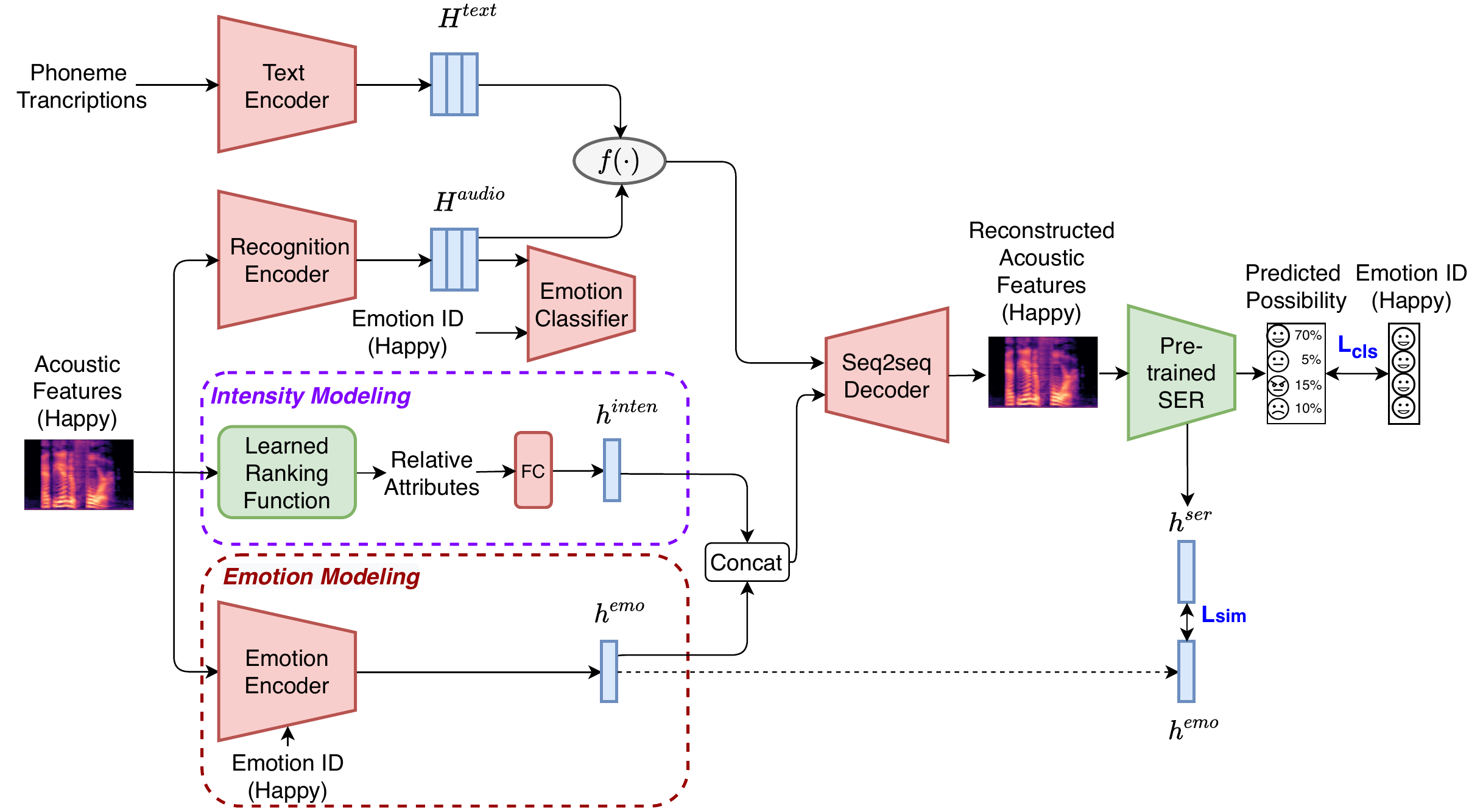}
    \caption{Overall training diagram of \textit{Emovox}, where emotion and its intensity are separately modelled.
     Two utterance-level perceptual losses from a pre-trained SER: 1) emotion similarity loss $L_{sim}$, and 2) emotion classification loss $L_{cls}$ are introduced to improve the emotional intelligibility at the utterance level. The red boxes represent the models that are involved in the training, while the green boxes are not.} 
    \label{fig:stage3}
\end{figure*}
\subsection{Seq2Seq Emotional Voice Conversion}
Human speech can be viewed as a combination of speech style, and linguistic content \cite{hirschberg2002communication,xu2011speech}. If the speech style that represents the emotion can be disentangled from the linguistic content, emotion conversion can be achieved by manipulating the speech style at run-time
while keeping the linguistic content and speaker identity unchanged \cite{gao2019nonparallel,schnell2021emocat}.

There are various ways to disentangle the speech elements. In \cite{zhang2019non}, text information and adversarial learning 
are used in a sequence-level autoencoder. This framework achieves strong disentanglement between linguistic and speaker representations and enables duration modelling for voice conversion. We adopt this framework in \textit{Emovox} to model emotion styles and intensity, as shown in Figure \ref{fig:stage3}. \textcolor{black}{To overcome the issues such as deletion and repetition with the Seq2Seq approach, we include a text input as the supervision signal to augment the linguistic embedding, which are shown effective in recent studies~\cite{zhang2019improving, zhang2019non, zhou21b_interspeech}.} 
 
Given the phoneme sequences and acoustic features as the input, the text encoder and the recognition encoder learn to predict the linguistic embedding from the text ($\mathbf{H}^{text}$) and the audio input ($\mathbf{H}^{audio}$), respectively. The emotion encoder learns the emotion representations from the speech, while the emotion classifier further eliminates the residual emotion information in the linguistic embedding $\mathbf{H}^{audio}$. The Seq2Seq decoder $Dec$  learns to reconstruct the acoustic features $\hat{\mathbf{A}}$ from the combination of the emotion embedding $\mathbf{h}^{emo}$, the intensity embedding $\mathbf{h}^{inten}$, and the linguistic embedding either from the text encoder: $\mathbf{H}^{text}$ or recognition encoder: $\mathbf{H}^{audio}$
as shown in~\eqref{eqn_Seq2Seq}.

\begin{equation}
    \hat{A} = Dec(\mathbf{h}^{emo}, \mathbf{h}^{inten}, f(epoch)),
    \label{eqn_Seq2Seq}
\end{equation}
where
\begin{equation}
    f(epoch) = \left\{ \begin{array}{rcl}
\mathbf{H}^{text} & \mbox{for}
& epoch \% 2=0 \\ 
\mathbf{H}^{audio} & \mbox{for} & epoch \% 2= 1
\end{array}\right.
\end{equation}

During the training, $\mathbf{H}^{text}$ and $\mathbf{H}^{audio}$ are taken by the decoder alternately, depending on whether the epoch number is odd or even. A contrastive loss is employed to ensure the similarity between $\mathbf{H}^{text}$ and $\mathbf{H}^{audio}$ as in~\cite{zhang2019non}. 
We believe the proposed \textit{Emovox} learns an effective disentanglement between linguistic and emotional elements and provides a straightforward way to model and control both emotion and its intensity, which will be discussed next. 

\subsection{Modelling Emotion and its Intensity}
To model emotion intensity, one of the difficulties is the lack of annotated intensity labels. 
Inspired by the idea of attribute \cite{ferrari2007learning} in computer vision, we regard emotion intensity as an attribute of the emotional speech. 
Combining the emotion representations with the intensity information allows the framework to jointly learn abundant emotion styles and intensity levels from any emotional speech database. 

\subsubsection{Formulation of Emotion Intensity using Relative attributes}
In computer vision, there are various ways \cite{mcfee2010metric,koch2015siamese} to model the relative difference between different data categories. Instead of predicting the presence of a specific attribute,
relative attributes \cite{parikh2011relative} offer more informative descriptions to unseen data, thus closer to detailed human supervision. Motivated by the success in various computer vision tasks \cite{kovashka2012whittlesearch,zhang2015robust,fan2013relative}, we believe that relative attributes bridge between the low-level features and high-level semantic meanings, which is appropriate for emotion intensity modelling.

Emotion intensity can be viewed as how well the emotion can be perceived in its type. Since the neutral speech does not contain any emotional variance, the emotion intensity of a  neutral utterance should be zero.
Therefore, we regard the emotion intensity as a relative difference between neutral speech and emotional speech. 
Emotion intensity can be represented by relative attributes learnt with a rich set of emotion-related acoustic features from each emotion pair.
The learning process of relative attributes can be formulated as a max-margin optimization problem as explained below:

Given a training set $T = \{\mathbf{x}_t\}$, where ${\mathbf{x}_t}$ is the acoustic features of the $t^{th}$ training sample, and $T = N \cup E$, where $N$ and $E$ are the neutral and emotional set respectively. We aim to learn a ranking function given as below:
\begin{equation}
    r(x_t) = \mathbf{W}\mathbf{x}_t, 
\end{equation}
where $\mathbf{W}$ is a weighting matrix indicating the emotion intensity. To learn the ranking function, we have to satisfy the following constraints:
\begin{align}
    \forall (a, b) \in O: \mathbf{W}\mathbf{x}_a > \mathbf{W}\mathbf{x}_b\\
    \forall (a, b) \in S: \mathbf{W}\mathbf{x}_a = \mathbf{W}\mathbf{x}_b, 
\end{align}
where $O$ and $S$ are the ordered and similar sets respectively. We pair an emotional sample of $E$ with a neutral sample from $N$ to form an ordered set $O$, where the emotion intensity of $E$ is higher than in that of $N$. We then randomly create pairs of neutral-neutral and emotional-emotional samples in the similar set $S$, where the emotion intensity of the pair is similar. The weighting matrix $\mathbf{W}$ is estimated by solving the following problem similar with that of a support vector machine \cite{chapelle2007training}:

\begin{align}
    \min_{\mathbf{W}} (\frac{1}{2} \parallel \mathbf{W} \parallel_2^2 + C(\sum \xi_{a,b}^2 + \sum \gamma_{a,b}^2))\\
    \text{ s.t. }           \mathbf{W}(\mathbf{x}_a - \mathbf{x}_b) \geq 1 - \xi_{a,b}; \forall(a,b) \in O\\
    |\mathbf{W}(\mathbf{x}_a-\mathbf{x}_b)|\leq \gamma_{a,b}; \forall(a,b) \in S\\
    \xi_{a,b} \leq 0; \gamma_{a,b}\leq 0, 
\end{align}
where $C$ is the trade-off between the margin and the size of slack variables $\xi_{a,b}$ and $\gamma_{a,b}$. 
Through Eq. (6) - (9), we learn a wide-margin ranking function that enforces the desired ordering on each training point. Once it is learnt, the relative ranking function can estimate the order 
of 
unseen data.
In practice, we learn a ranking function for each emotion category.
As shown in Figure \ref{fig:stage3},
the learnt ranking function predicts a relative attribute normalized to $[0,1]$ for each sample in the training set. A larger value of relative attribute represents a stronger intensity of an emotion.

\subsubsection{Modelling Emotion Styles and its Intensity}
As shown in Figure \ref{fig:stage3}, we obtain the relative attribute from the learnt ranking function, which passes through a fully connected layer to derive an intensity embedding. The emotion encoder learns to generate the emotion embedding from the input speech features.
The Seq2Seq decoder combines a linguistic embedding sequence, an emotion and an intensity embedding to reconstruct the acoustic features of the emotional speech.

During the training process, \textit{Emovox} jointly learns the emotion style and its intensity from the speech samples that are referred to as \textit{emotion training} hereafter. With the explicit intensity modelling, we are able to manipulate the level of intensity at run-time for intensity control. The intended emotion intensity can be predicted from the reference or given manually at run-time. 
\textcolor{black}{In theory, \textit{Emovox} may perform both emotional text-to-speech and emotional voice conversion. In this paper, the text encoder is not used at run-time since we are only interested in voice conversion.} 

As shown in Figure \ref{fig:conversion}, we first use the emotion encoder to generate the emotion embeddings from a set of reference utterances belonging to the same emotional category. Next, we use the averaged reference emotion embedding to represent an emotion category. Finally, the recognition encoder derives a linguistic embedding sequence from the source speech utterance at run-time. By assigning an intended emotion category and a level of emotion intensity, the Seq2Seq decoder generates the emotional speech of the same content as the source but with the target emotion style at an appropriate intensity.

\subsection{Model Pre-training}

 During training, a large amount of emotional speech is always required to achieve robust attention alignment and deliver high emotional intelligibility in a Seq2Seq model \cite{lee2020memory}. To reduce the reliance on emotional speech, we propose two pre-training strategies,
1) style pre-training with a large TTS corpus and 2) emotion supervision training with a SER.

\subsubsection{Style Pre-training with a Multi-Speaker TTS Corpus}
It is known that speech style contains speaker-dependent elements related to speaker characteristics, called speaker style. 
Speaker style is exhibited in most TTS corpora containing multi-speaker speech data. Unlike emotional speech databases, there are abundant speech databases for TTS \cite{kominek2004cmu,ljspeech17,veaux2016vctk} with a neutral tone, which allows us to build a multi-speaker Seq2Seq TTS framework, and train a network to disentangle speaker style from the linguistic content.   We call this stage ''style pre-training''. 

During the style pre-training, the style encoder learns abundant speaker styles through a multi-speaker TTS corpus while excluding the linguistic information from the acoustic features. As a result, even though the style encoder does not learn to encode any specific emotion style during training, it learns to discriminate different emotion styles during emotion training, as shown in Figure \ref{fig:embedding}(a). We, therefore, use the style encoder trained on a TTS corpus as the pre-trained model for an emotion encoder. 
\begin{figure}[t]
    \centering
    \includegraphics[width=0.5\textwidth]{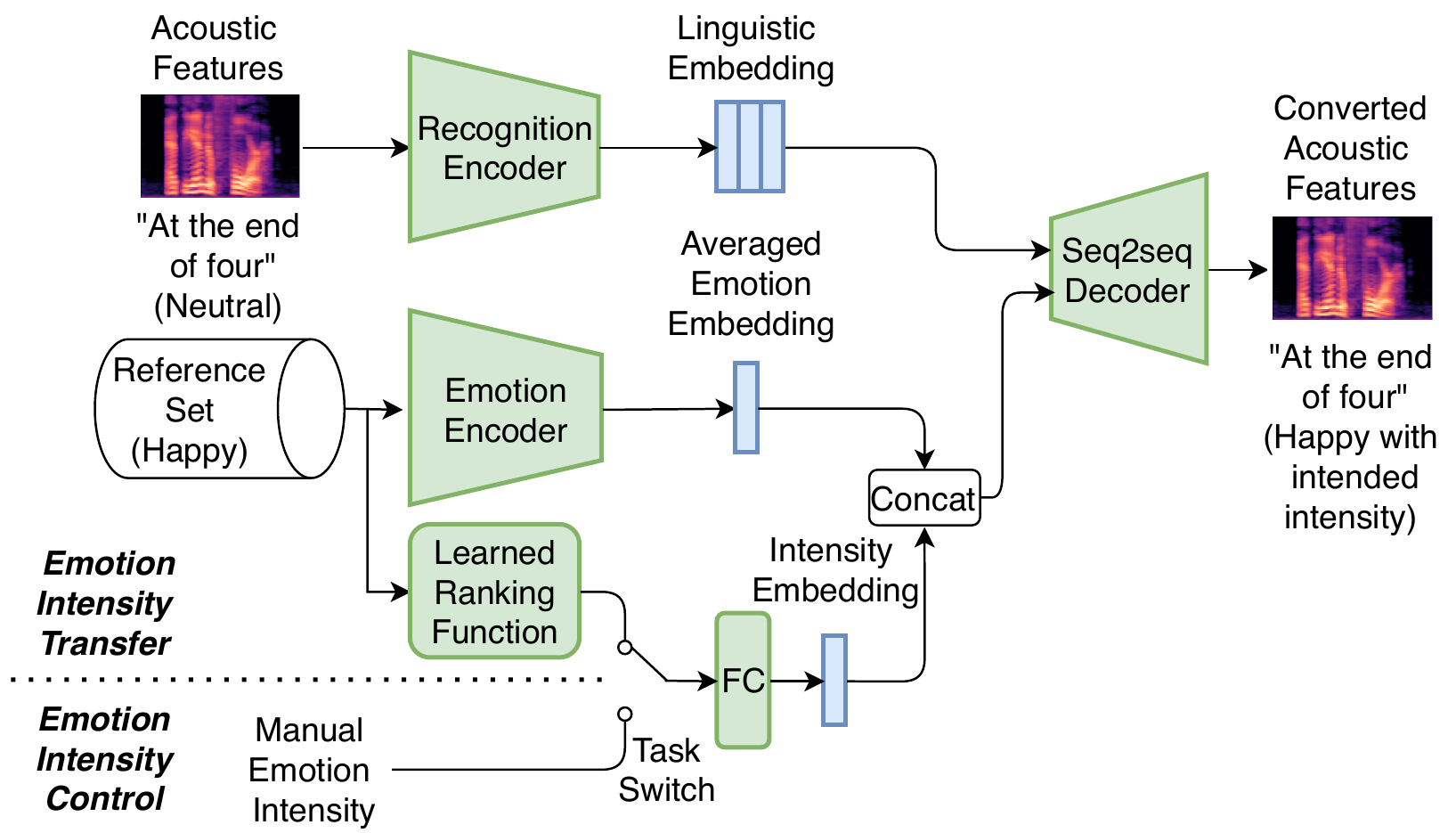}
    \caption{An illustration of the run-time conversion phase.
    By combining a source linguistic embedding sequence, an averaged reference emotion embedding, and an intensity embedding, the Seq2Seq decoder generates the acoustic features with the reference emotion type and the manually defined intended intensity.}
    \label{fig:conversion}
\end{figure}
\begin{figure*}[t]
    \centering
    \includegraphics[width=1\textwidth]{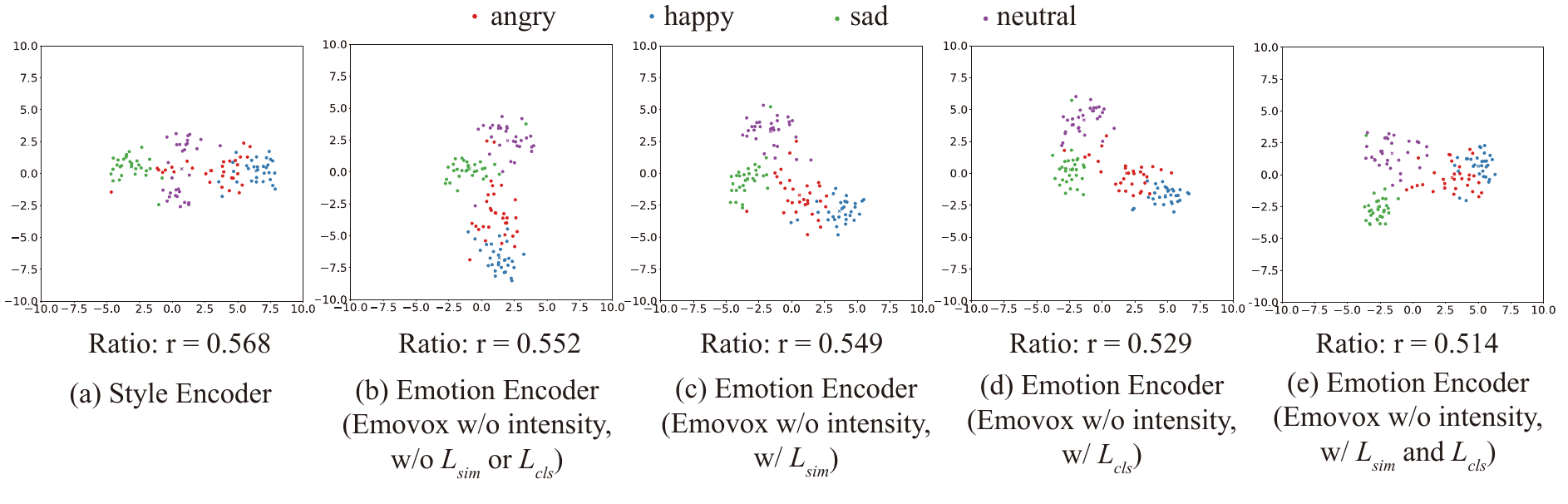}
    \caption{
    The distributions of emotion embeddings resulting from encoders of different training schemes: (a) the pre-trained style encoder, (b) the emotion encoder without $L_{sim}$ or $L_{cls}$, (c) the emotion encoder only with $L_{cls}$,  (d) the emotion encoder only with $L_{sim}$, and (e) the emotion encoder with both $L_{sim}$ and $L_{cls}$.   A smaller value of ratio $r$ indicates a better clustering performance.} 
    \label{fig:embedding}
\end{figure*}
\subsubsection{Modelling Emotion with Perceptual Loss}

We would like the converted emotional speech to be perceived with the intended emotion category. However, this is not easily achieved, especially with 
 a limited emotional training data, for several reasons: 1) The pre-trained emotion decoder in Figure \ref{fig:stage3} is not explicitly trained for characterisation of emotions, 
and 2) frame-level style reconstruction loss is not always consistent with human perception because it does not capture speech's prosodic and temporal patterns.

Following the success of perceptual loss in speech synthesis~\cite{oord2018parallel}, we introduce a perceptual loss as the emotion supervision in the training process, as shown in Figure \ref{fig:stage3}. We first use a pre-trained SER to predict the emotion category from the reconstructed acoustic features. 
We then calculate two perceptual loss functions: 1) emotion classification loss $L_{cls}$, and 2) emotion embedding similarity loss $L_{sim}$. We incorporate these two loss functions into the training and update all the trainable modules. \textcolor{black}{For detail of SER pre-training, readers are referred to \cite{ser}.}  

The emotion classification loss $L_{cls}$ is introduced to ensure the perceptual similarity between the reconstructed acoustic features and the intended emotion category at the utterance level,
\begin{equation}
    L_{cls} =  \mathrm{CE}(\mathbf{l}, \hat{\mathbf{p}}),
\end{equation}
where $\mathbf{l}$ is the target one-hot emotion label, $\hat{\mathbf{p}}$ is the predicted emotion probabilities at the utterance level, and $\mathrm{CE(\cdot)}$ denotes the cross-entropy loss function. 

{The pre-trained SER is considered text-independent. To ensure that the emotion encoder characterizes emotions independent of linguistic content, we introduce an emotion style descriptor, derived from the pre-trained SER} 
\textcolor{black}{~\cite{zhou2021seen, liu2021expressive}, as a learning objective for emotion encoder with a loss function $L_{sim}$, 
between the emotion encoder output, i.\,e., emotion embedding, and the emotion style descriptor, as illustrated in Figure \ref{fig:stage3}. }

\begin{equation}
    L_{sim} = \sqrt{ \frac{1}{D}\sum_{d=1}^{D}(\mathbf{h}^{emo}_d - \mathbf{h}^{ser}_d)^2},
\end{equation}
where $\mathbf{h}^{emo}_d$ is the emotion embedding derived from the emotion encoder of $D$ dimensions, and $\mathbf{h}^{ser}_d$ is the emotion style descriptor derived right before the last projection layer of the pre-trained SER.

\subsubsection{Effect of Perceptual Loss}
\label{sec: perceptual}

To validate the effectiveness of two perceptual loss functions, we evaluate the emotion-discriminative ability of the emotion encoder with an ablation study. We believe that the emotion encoder demonstrates a better performance by producing more discriminative emotional representations. 

We use the t-SNE algorithm \cite{maaten2008visualizing} to visualize the emotion embeddings in a two-dimensional plane, as shown in Figure \ref{fig:embedding}(e). It is observed that emotion embeddings form different emotion clusters in terms of the feature distribution. To get a more intuitive understanding of the clustering performance, we consider performing a clustering evaluation to evaluate the discriminability of the emotion embeddings.

The typical objective function of clustering formalizes the goal of attaining high intra-cluster similarity, and low inter-cluster similarity \cite{maulik2002performance,sarle1990algorithms}. There are studies to use different measurements for the quality of a clustering \cite{zhao2002evaluation,larose2014discovering,liu2010understanding,rosenberg2007v}. 
Our study considers a simplified and effective solution for clustering evaluation.
We first compute a centroid for each of $K$ emotion classes, $\mathbf{c}_i$, $i \in [1,K]$ by taking the average of all $N_i$ embeddings $\mathbf{e}$ in class $i$ as follows \cite{kwon2019effective}:
\begin{equation}
    \mathbf{c}_i = \frac{1}{N_i} \sum_{\mathbf{e}\in E_i} \mathbf{e}, 
\end{equation}
where $E_i$ is the set of embeddings in class $i$. 
We then calculate the inter-class distance $dist_{inter}$ by computing the Euclidean distance between each embedding $\mathbf{e} \in E_i$ and the other embedding centres $\mathbf{c}_{j; j\neq i}$ as follows:
\begin{equation}
    dist_{inter} = \frac{1}{K(K-1)}\sum_{i=1}^K \frac{1}{N_i} \sum_{\mathbf{e}\in E_i} \sum_{j; j\neq i} \sqrt{(\mathbf{e}-\mathbf{c}_j)^2}
\end{equation}
and intra-class distance $dist_{intra}$ as follows:
\begin{equation}
    dist_{intra} = \frac{1}{K}\sum_{i=1}^K\frac{1}{N_i} \sum_{\mathbf{e}\in E_i} \sqrt{(\mathbf{e} - \mathbf{c}_i)^2}.
\end{equation}
A clustering ratio $r$ is calculated from the ratio of intra-class distance $dist_{intra}$ and inter-class distance $dist_{inter}$ as follows:
\begin{equation}
    r = \frac{dist_{intra}}{dist_{inter}}.
\end{equation}
A lower value of ratio $r$ represents a better clustering effect of emotion embeddings.

We perform an ablation experiment on the ESD evaluation dataset~\cite{zhou2021emotional}. We visualize the distribution of emotion embeddings and report the clustering ratios in Figure \ref{fig:embedding}. As the style encoder is pre-trained without the emotion intensity mechanism, we report the results of \textit{Emovox} without intensity control for a fair comparison, which is denoted as \textit{Emovox w/o intensity}. We first observe that \textit{Emovox w/o intensity} always achieves a better clustering performance than the style encoder in Figure \ref{fig:embedding}.
From Figure \ref{fig:embedding}(b) - (d), it is observed that both loss functions $L_{cls}$ and $L_{sim}$ contribute to a lower $r$, which suggests a better clustering performance. From Figure \ref{fig:embedding}(e), we further observe a more distinct separation between the emotions with different energy (such as neutral, sad vs angry, or happy). With both $L_{cls}$ and $L_{sim}$, we obtain the lowest clustering ratio at $0.514$. It shows that these two losses can help the emotion encoder to generate more discriminative emotional representations.

\section{Experiments}
\label{sec: exp}
In this section, we report our experimental settings. For all the experiments, we conduct emotion conversion from neutral to angry, neutral to happy, and neutral to sad, which we denote as \textit{Neu-Ang}, \textit{Neu-Hap}, and \textit{Neu-Sad}, respectively. We have made the source codes and speech samples available to the public\footnote{\textbf{Codes \& Speech Samples}: \href{https://kunzhou9646.github.io/Emovox_demo/}{\nolinkurl{https://kunzhou9646.github.io/Emovox\_demo/}}}. We encourage readers to listen to the speech samples to understand this work. 

\subsection{Reference Methods and Setups}
\label{sec: reference}
We implement 3 state-of-the-art emotional voice conversion methods as the reference baselines, that are summarized as follows:
\begin{itemize}
    \item \textbf{CycleGAN-EVC} \cite{Zhou2020} \textit{(baseline)}: CycleGAN-based emotional voice conversion with WORLD vocoder \cite{morise2016world}, where the fundamental frequency (F0) is analyzed with continuous wavelet transform;
    \item \textbf{StarGAN-EVC} \cite{rizos2020stargan} \textit{(baseline)}: StarGAN-based emotional voice conversion with WORLD vocoder \cite{morise2016world}; 
    \item \textbf{Seq2Seq-EVC} \cite{zhou21b_interspeech} \textit{(baseline)}: Sequence-to-sequence emotional voice conversion with a Parallel WaveGAN vocoder \cite{yamamoto2020parallel};
    \item \textbf{Emovox} \textit{(proposed)}: Our proposed sequence-to-sequence emotional voice conversion framework with a Parallel WaveGAN vocoder \cite{yamamoto2020parallel} shown in Figure 3. 
\end{itemize}
{Note that emotion intensity control is only available with \textit{Emovox}. For a fair comparison among the methods, we obtain an intensity value for \textit{Emovox}, by passing a reference set of speech data through the learnt ranking function, as shown in Figure \ref{fig:conversion} ("Emotion Intensity Transfer") .} Besides, none of these frameworks require any parallel training data or frame alignment procedures. 

{For a contrastive study,  we replace the  intensity control module in \textit{Emovox} with two other competing intensity control methods: Scaling Factor and Attention Weights through comprehensive experiments.}

\begin{itemize}
    \item \textbf{Emovox w/ Scaling Factor} \textit{(proposed)}: where the emotion embedding is multiplied by a scaling factor \cite{choi2021sequence};
    
    \item\textbf{Emovox w/ Attention Weights} \textit{(proposed)}: where the attention weight vector obtained from a pre-trained SER is used to represent the intensity \cite{schnell11improving};
    \item \textbf{Emovox w/ Relative Attributes} \textit{(proposed)}: our proposed method with relative attributes as described in Section 3;
\end{itemize}
To summarize, we do emotion intensity transfer to compare Emovox with the baselines (i.\,e., CycleGAN-EVC, StarGAN-EVC and Seq2Seq-EVC) and emotion intensity control to compare it with other emotion intensity control methods (i.\,e., Emovox w/ scaling factor, Emovox w/ attention weights).

\begin{table*}[t]
\centering
\caption{A comparison of the MCD and the DDUR results of different methods for three emotion conversion pairs.}
\begin{tabular}{c|ccc|ccc}
\hline
\multirow{2}{*}{Framework} & \multicolumn{3}{c|}{MCD {[}dB{]}} & \multicolumn{3}{c}{DDUR {[}s{]}} \\ \cline{2-7} 
                           & Neu-Ang   & Neu-Hap   & Neu-Sad   & Neu-Ang   & Neu-Hap   & Neu-Sad   \\ \hline
Zero Effort              & 6.47      & 6.64      & 6.22      & 0.36      & 0.26      & 0.46      \\ 
CycleGAN-EVC               & 4.57      & 4.46      & 4.32      & -      & -      & -     \\ 
StarGAN-EVC                & 4.43      & 4.25      & 4.31      & -     & -      & -      \\ 
Seq2Seq-EVC                & 4.29      & 4.16          & \textbf{4.23}          & 0.28      &   0.20        &  \textbf{0.27}         \\
Emovox (w/o style pre-training)                   & 5.36          &  5.32         & 5.42          &    0.79       &  0.80         & 0.92 \\
Emovox                   & \textbf{4.13}          &  \textbf{4.15}         & 4.25          &    \textbf{0.24}       &  \textbf{0.17}         & 0.31          \\ \hline
\end{tabular}
\begin{tablenotes}
\centering
\footnotesize
\item Note: DDUR results of CycleGAN-EVC and StarGAN-EVC are not reported, as they cannot modify the speech duration.
\end{tablenotes}
\label{tab:objective1}
\end{table*}

\subsection{Experimental Setup}

We extract 80-dimensional Mel-spectrograms every $12.5$\,ms with a frame size of $50$\,ms for short-time Fourier transform (STFT). We then take the logarithm of the Mel-spectrograms to serve as the acoustic features. We convert text to phoneme with the Festival \cite{black2001festival} G2P tool to serve as the input to the text encoder. 

We use the Adam optimizer \cite{kingmaadam} and set the batch size to $64$ and $16$ for style pre-training and emotion training, respectively. We set the learning rate to $0.001$ for style pre-training and halve it every seven epochs during the emotion training. We set the weight decay to $0.0001$, and the weighting factors of the emotion classification loss $L_{cls}$ and the emotion similarity loss $L_{sim}$ to 1. 

\subsubsection{Recognition-Synthesis Structure}
\textit{Emovox} has a recognition-synthesis structure similar to that of \cite{zhang2019non,zhang20d_interspeech}.
The Seq2Seq recognition encoder consists of an encoder which is a 2-layer 256-cell BLSTM, and a decoder which is a 1-layer 512-cell LSTM with an attention layer followed by an FC layer with an output channel of $512$.
Our text encoder is a 3-layer 1D CNN with a kernel size of $5$ and the channel number of $512$, followed by 1-layer of 256-cell BLSTM and an FC layer with an output channel number of $512$. The Seq2Seq decoder has the same model architecture as that of Tacotron \cite{wang2017tacotron}.
The style encoder is a 2-layer of 128-cell BLSTM followed by an FC layer with an output channel number of $128$, which has been used in previous studies on voice conversion \cite{zhang2019non} and emotional voice conversion \cite{zhou21b_interspeech}. The classifier is a 4-layer net of FC with the channel numbers of $\{$512, 512, 512, 99$\}$. 
\subsubsection{Relative Emotion Intensity Ranking} We follow an open-source implementation\footnote{\url{https://github.com/chaitanya100100/Relative-Attributes-Zero-Shot-Learning}} to train the relative ranking function for emotion intensity. We extract 384-dimensional acoustic features with openSMILE \cite{eyben2010opensmile} including zero-crossing rate, frame energy, pitch frequency, Mel-frequency cepstral coefficient (MFCC), and etc. 
These acoustic features are used in the Interspeech Emotion Challenge \cite{schuller2009interspeech}. We anticipate that these acoustic features can capture the subtle emotion intensity variations in speech.
For each emotion category, we train a relative ranking function using neutral and emotional utterances.

 \subsubsection{Speech Emotion Recognizer}We  train a speech emotion recognizer following a publicly available implementation \cite{ser}. The SER includes: 1) 3 TimeDistributed two-dimensional (2-D) convolutional neural network (CNN) layers, 2) a DBLSTM layer, 3) an attention layer, and 4) a linear projection layer. The TimeDistributed 2D CNN layers and the DBLSTM layer summarize the temporal information into a fixed-length latent representation. The attention layer further preserves the effective emotional information while reducing the influence of emotion-irrelevant factors and producing discriminative utterance-level features for emotion prediction. The linear projection layer predicts the emotion class possibility from the utterance-level emotional features. We perform data augmentation by adding white Gaussian noise to improve the robustness of SER (\cite{heracleous2017speech,tiwari2020multi,abbaschian2021deep,muthusamy2015improved}).

\subsubsection{Data Preparation and Emotion Training}
We first perform style pre-training on the VCTK Corpus \cite{veaux2016vctk}, where we use $99$ speakers for pre-training. The total duration of pre-training speech data is about 30 hours. 
For SER training and emotion training,
we randomly choose one male speaker from the ESD database\footnote{ \href{hhttps://hltsingapore.github.io/ESD/download.html}{\nolinkurl{https://hltsingapore.github.io/ESD/download.html}}} to conduct all the experiments in the same way as in \cite{zhou21b_interspeech}. 

{We follow the data partition protocol given in the ESD database. For each emotion, we use 300 utterances for emotion training and 20 utterances as the evaluation set. We use 30 utterances to form a reference set to generate the reference emotion embeddings for each emotion category at run-time.}  The total speech duration of emotional training data is around \textit{50 minutes} (about 12 minutes for each emotion), which is very limited in the context of Seq2Seq training.

In the emotion training, we first initialize all the modules with the weights learnt from style pre-training, where the style encoder and style classifier act as the emotion encoder and emotion classifier, respectively. We then randomly initialize the last projection layer of the emotion encoder and emotion classifier. The output channel numbers of the emotion encoder and the emotion classifier are set to $64$ and $4$, respectively. A learnt ranking function predicts a relative attribute and then is passed through an FC layer 
with the output channel size of $64$ to obtain the intensity embedding. We then concatenate the emotion and intensity embedding to feed into the Seq2Seq decoder. The waveform is reconstructed from the converted Mel-spectrograms using Parallel WaveGAN. We use a public version of Parallel WaveGAN\footnote{https://github.com/kan-bayashi/ParallelWaveGAN}, and train it with the ESD database.

\subsection{Objective Evaluation}

We first conduct an objective evaluation to assess the system performance using Mel-cepstral Distortion (MCD) and Differences of Duration (DDUR) as the evaluation metrics.

\subsubsection{Mel-cepstral Distortion (MCD)}   MCD \cite{kubichek1993mel}  is calculated between the converted and the target Mel-cepstral coefficients (MCEPs), i.\,e.,  $\hat{\mathbf{y}}=\{\hat{\mathbf{y}}_m\}$ and ${\mathbf{y}}=\{\mathbf{{y}}_m\}$,
\begin{equation}
    \text{MCD [dB]} = \frac{10\sqrt{2}}{\ln 10}\frac{1}{M}\sqrt{\sum_{m=1}^M(\mathbf{y_{m}} - \hat{\mathbf{y}}_{m})^2}, 
\end{equation}
where $M$ represents the dimension of the MCEPs. \emph{A lower value of MCD indicates a smaller spectral distortion, and thus a better performance.}
Note that, in the Seq2Seq-EVC  and \textit{Emovox} models, we adopt Mel-spectrograms as the acoustic features. Therefore, we calculate MCEPs separately from the speech waveform.

\subsubsection{Differences of Duration (DDUR)}
To evaluate the distortion in terms of duration,  we compute the average differences between the duration of the converted and the target utterances over the voiced parts (DDUR), which is widely used in voice conversion studies \cite{zhang2019non,zhang2019sequence,zhou21b_interspeech},
\begin{equation}
\text{DDUR [s]} = |Z - \hat{Z}|,
\end{equation}
where $Z$ and $\hat{Z}$ represent the duration of the reference utterance and the converted utterance, respectively.
\emph{A lower value of DDUR represents a better performance in terms of duration conversion.}

\subsection{Subjective Evaluation}
\label{sec: subj}

We adopt two subjective metrics: 1) a mean opinion score (MOS) test for emotion similarity evaluation, and 2) a best-worst scaling (BWS) test to evaluate speech quality, emotion intensity, and emotion similarity. 18 subjects participated in all the listening tests. These 18 subjects (12 male and 6 female) are native Chinese speakers and proficient in English. Their age range is between 20-30. All the subjects are required to listen with headphones and replay each sample 2-3 times. A detailed introduction about the judging criteria is given before the tests. 
\begin{figure}[t]
    \centering
    \includegraphics[width=0.5\textwidth]{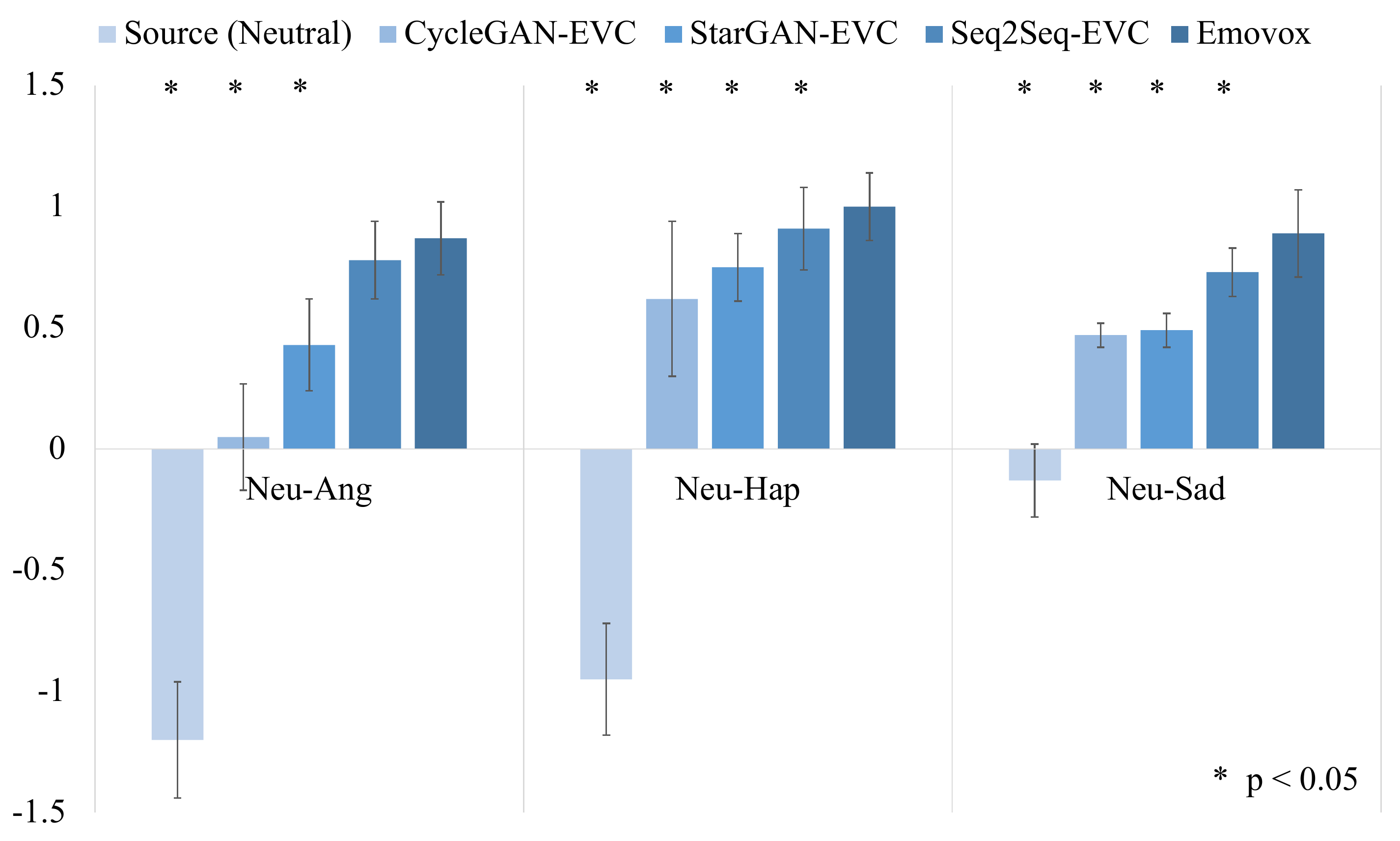}
    \caption{Mean Opinion Score (MOS) test with 95\,\% confidence interval to evaluate the emotion similarity with the reference, where listeners are asked to score each sample in a scale from -2 to +2 (-2: absolutely different; -1: different; 0: cannot tell; +1: similar; +2: absolutely similar). Marker * indicates p $<$ 0.05 for paired t-test scores (pairs between Emovox and the others).}
    \label{fig:mos}
\end{figure}
\subsubsection{Mean Opinion Score (MOS) Test}
\label{sec: mos}
We conduct a mean opinion score (MOS) \cite{streijl2016mean} test to evaluate the emotion similarity. All participants are asked to listen to the reference target speech first and then score the speech samples for emotion similarity to the reference target speech. 
A higher score represents a higher similarity with the target emotion, and indicates a better emotion conversion performance. We randomly select 10 utterances from the evaluation set. Each subject listens to 120 converted utterances in total (120 = 10 x 4 (\# of frameworks) x 3 (\# of emotion pairs)). 
\subsubsection{Best-Worst Scaling (BWS) Test}
\label{sec: bws}
We also conduct a best-worst scaling (BWS) \cite{kiritchenko2017best} test to evaluate:
\begin{enumerate}
\item \textbf{Speech Quality}: where all the listeners are asked to choose the best and the worst sample in terms of the speech quality, which covers two aspects: a) how the linguistic and speaker identity is preserved, and b) the naturalness of the speech;
\item \textbf{Emotion Intensity}: where all the listeners are asked to choose the most and the least expressive one in terms of the emotion expression;
\item \textbf{Emotion Similarity}: where all the listeners are asked to choose the best and the worst one in terms of the emotion similarity with the reference.
\end{enumerate}

We randomly select 5 utterances from the evaluation set to perform the BWS tests. 
We first evaluate the performance of different intensity control methods in terms of speech quality and intensity control.
Each subject listens to 135 converted utterances (135 = 5 x 3 (\# of frameworks) x 3 (\# of intensities) x 3 (\# of emotion pairs)). We further conduct an ablation study with \textit{Emovox}, where each subject listens to 60 converted utterances in total to evaluate the emotion similarity with the reference (60 = 5 x 4 (\# of frameworks) x 3 (\# of emotion pairs)).

\begin{figure}[t]
\centering
\subfloat[Source Neutral (Intensity = 0)]
{
\begin{minipage}[c]{1\linewidth}
\centering
\includegraphics[width=0.95\textwidth]{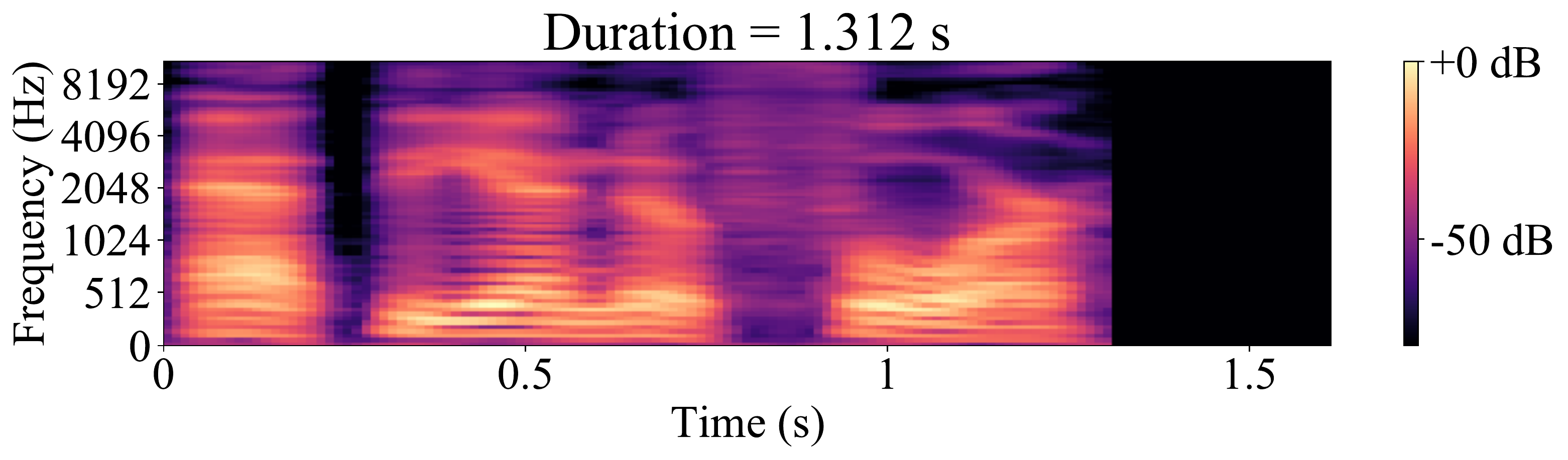}
\end{minipage}%
}
\newline
\subfloat[Reference Sad (Intensity = 0.611)]
{
\begin{minipage}[c]{1\linewidth}
\centering
\includegraphics[width=0.95\textwidth]{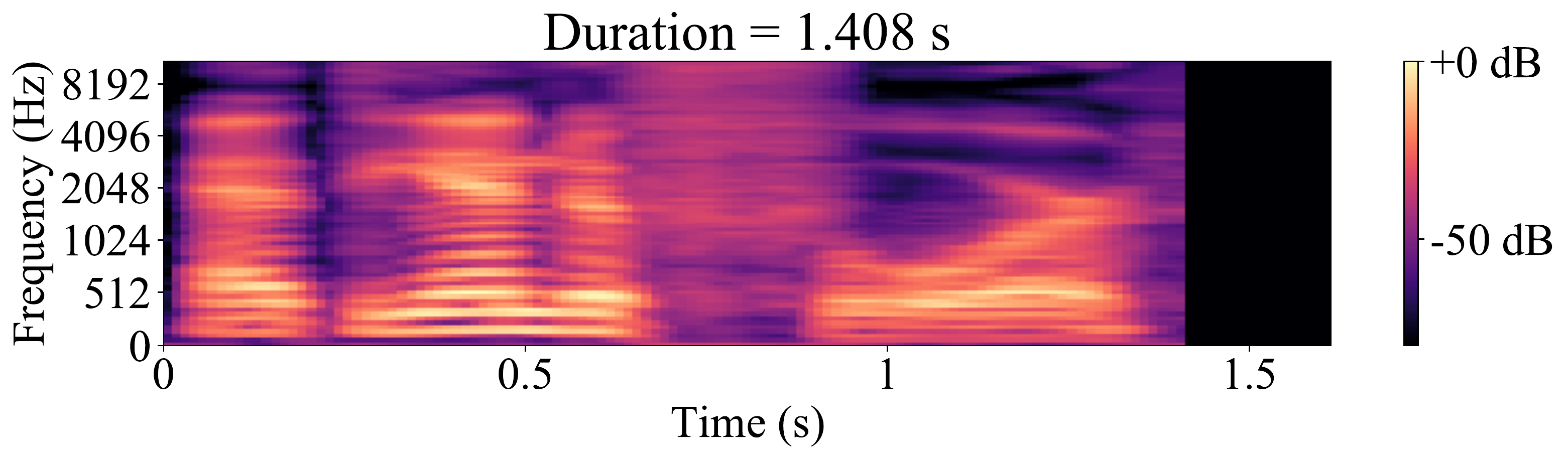}
\end{minipage}
}
\newline
\subfloat[Converted Sad (Intensity = 0.1) ]
{
\begin{minipage}[c]{1\linewidth}
\centering
\includegraphics[width=0.95\textwidth]{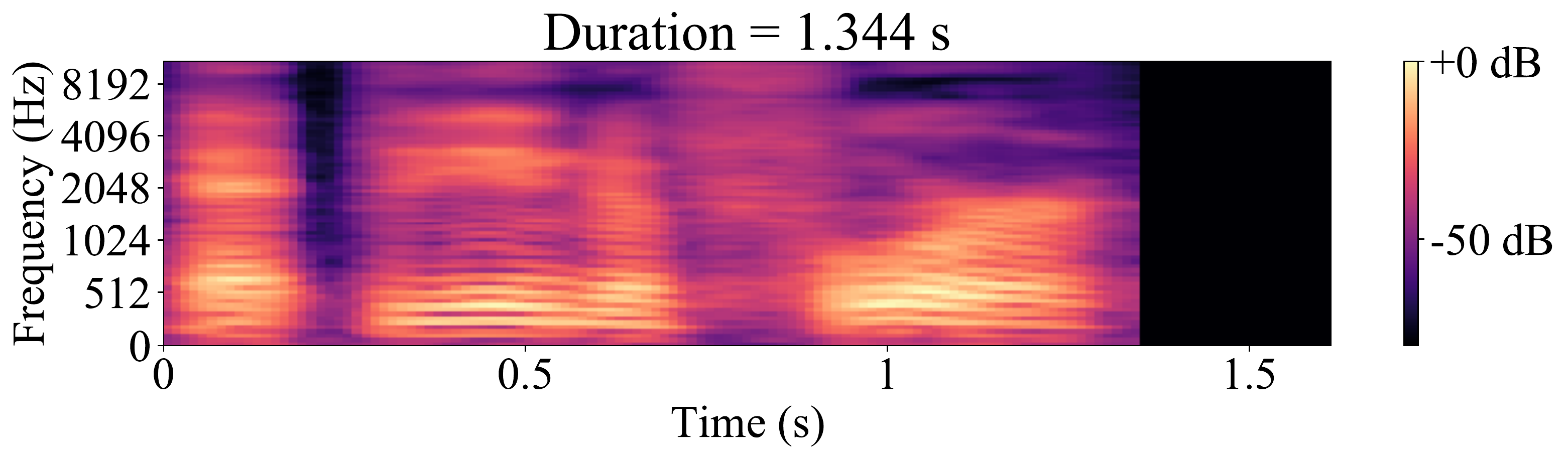}
\end{minipage}
}
\newline
\subfloat[Converted Sad (Intensity = 0.5) ]
{
\begin{minipage}[c]{1\linewidth}
\centering
\includegraphics[width=0.95\textwidth]{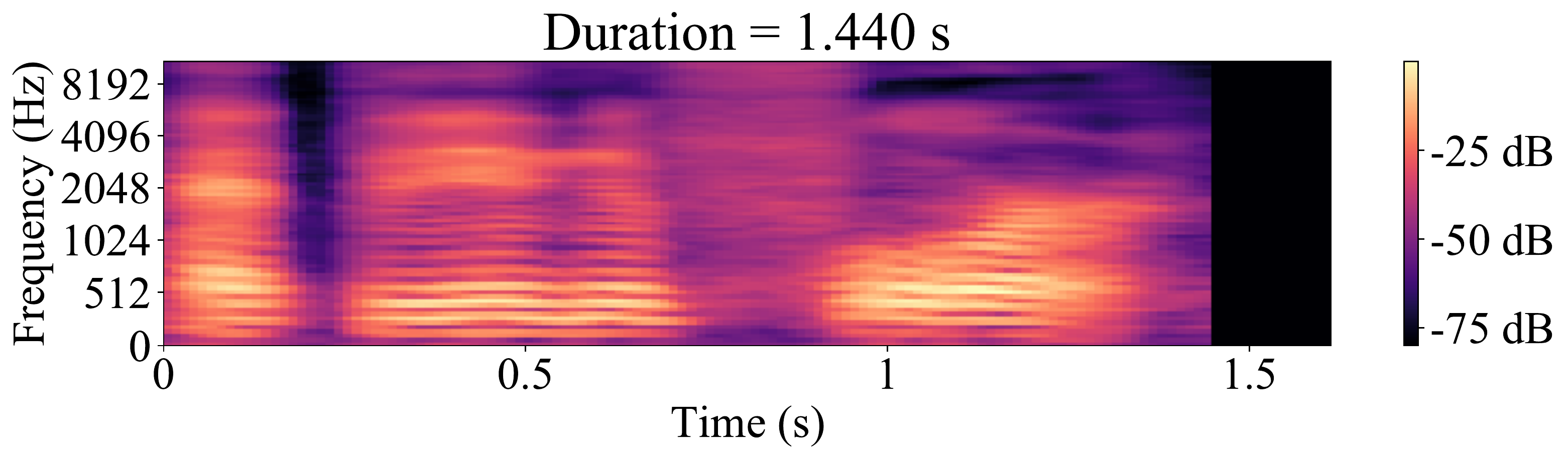}
\end{minipage}
}
\newline
\subfloat[Converted Sad (Intensity = 0.9) ]
{
\begin{minipage}[c]{1\linewidth}
\centering
\includegraphics[width=0.95\textwidth]{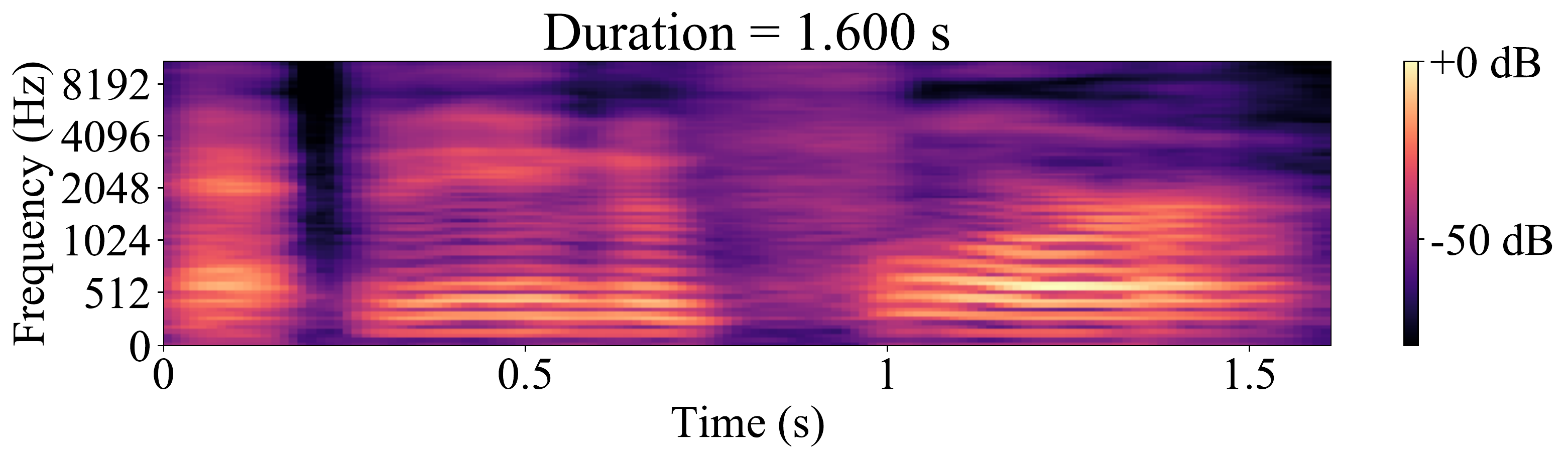}
\end{minipage}
}
\centering
\caption{Visualization of Mel-spectrograms from source neutral, reference sad, and converted sad utterances at three intensity values, i.\,e., 0.1, 0.5, 0.9, with the same speaking content (''At the end of four''). A greater intensity value represents a more emotional expression.}
\label{fig:ddur}
\end{figure}
\section{Results}
\label{sec: results}
In this section, we report our experimental results. We first compare the performance of \textit{Emovox} with that of the baselines using objective and subjective evaluations in Section \ref{sec: baseline}. We then evaluate the proposed emotion intensity control method through the comparison with other control methods in Section \ref{sec: intensity control}. While comparing with the baselines, we use different training data settings in Section \ref{sec: size}. Lastly, we study the contributions of the training strategies using ablation experiments in Section \ref{sec: ablation}.
\begin{figure*}[t]
\centering
\subfloat[Converted Angry]
{
\begin{minipage}[c]{0.33\linewidth}
\centering
\includegraphics[width=1\textwidth]{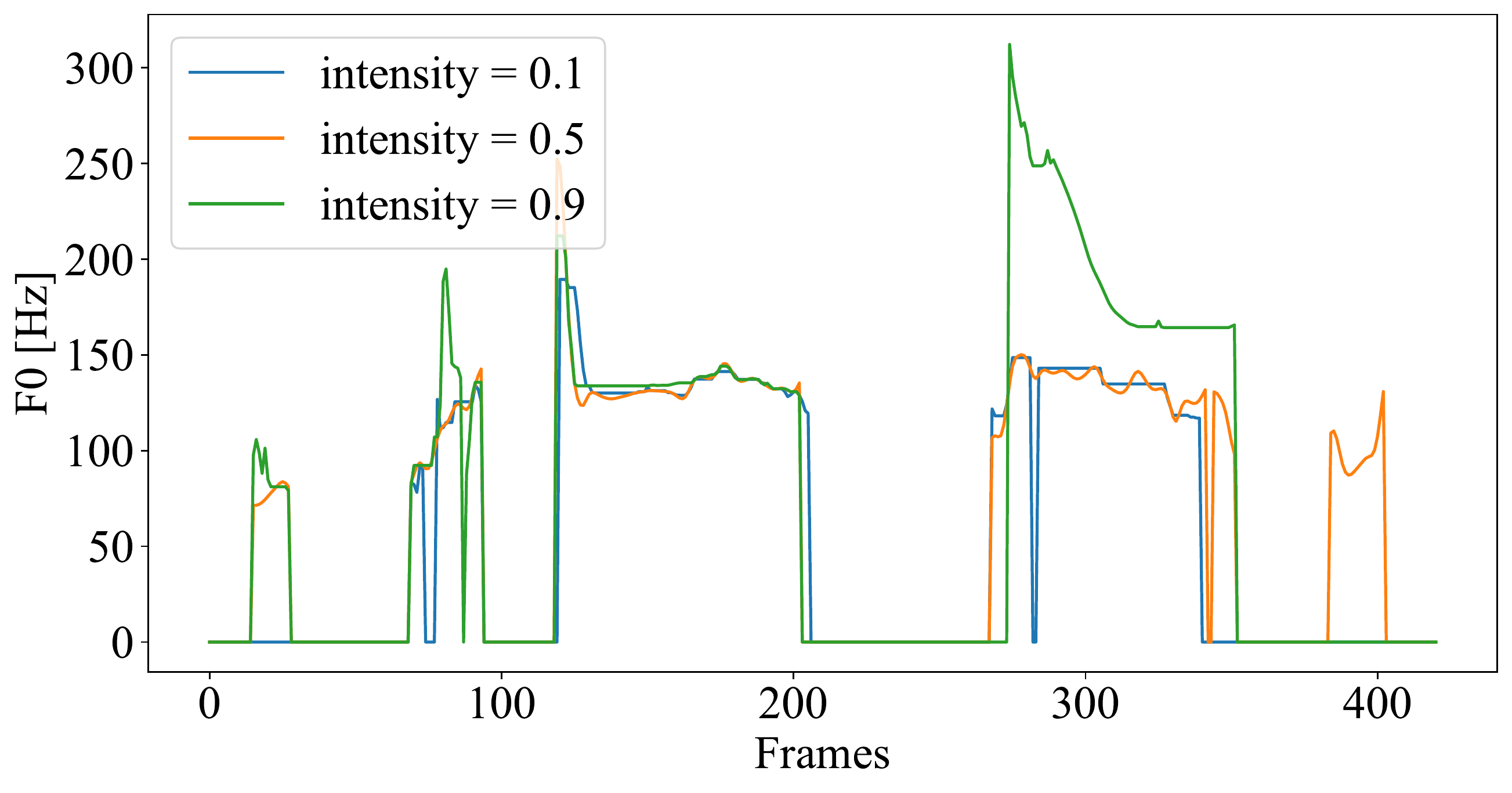}
\end{minipage}%
}
\subfloat[Converted Happy]
{
\begin{minipage}[c]{0.33\linewidth}
\centering
\includegraphics[width=1\textwidth]{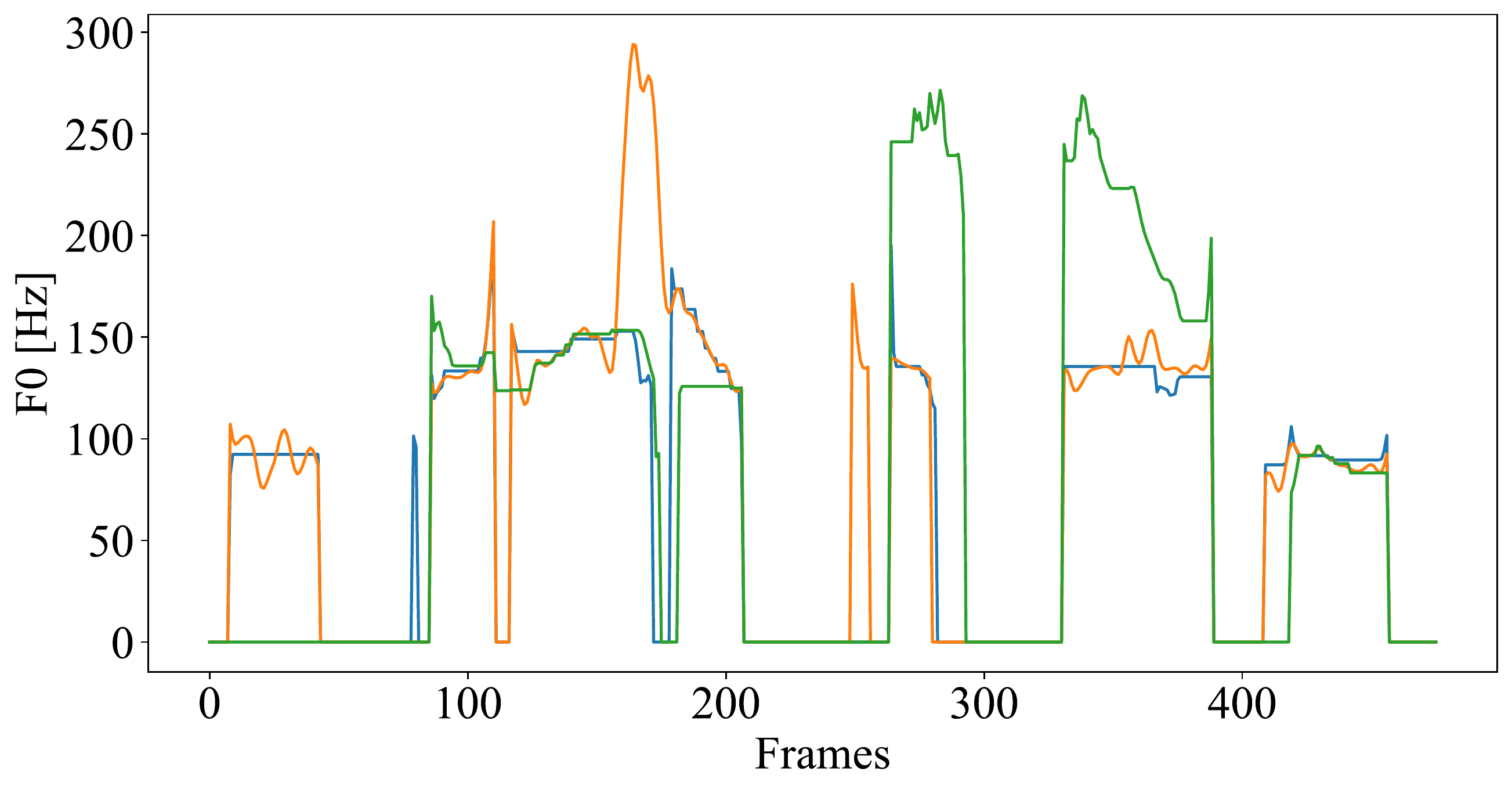}
\end{minipage}
}
\subfloat[Converted Sad]
{
\begin{minipage}[c]{0.33\linewidth}
\centering
\includegraphics[width=1\textwidth]{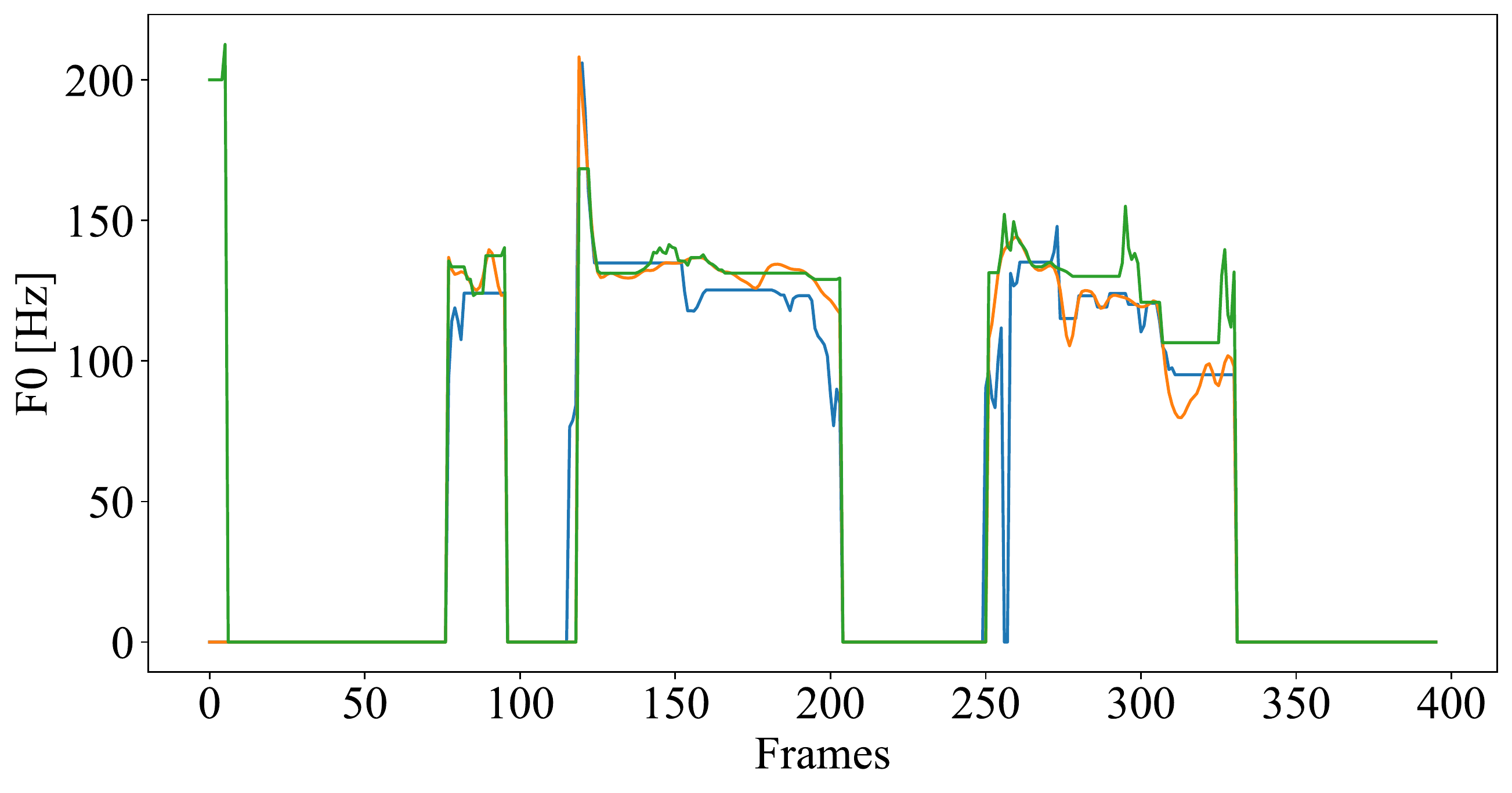}
\end{minipage}
}
\centering
\caption{A comparison of the pitch contour from the emotional utterances converted by \textit{Emovox} with three different emotion intensities (0.1, 0.5 and 0.9). }
\label{fig:pitch}
\end{figure*}

\begin{figure*}[t]
\centering
\subfloat[Converted Angry]
{
\begin{minipage}[c]{0.33\linewidth}
\centering
\includegraphics[width=1\textwidth]{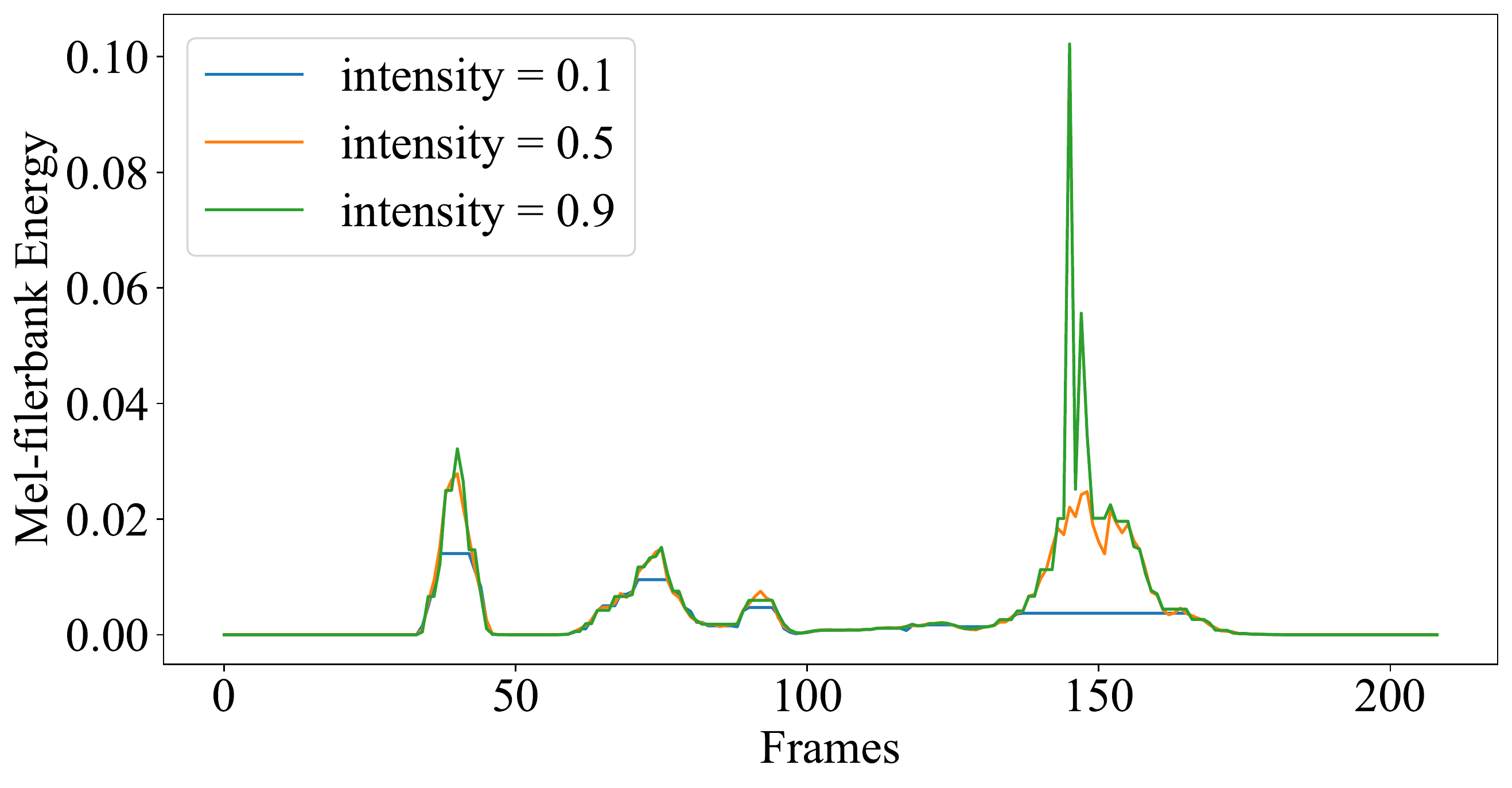}
\end{minipage}%
}
\subfloat[Converted Happy]
{
\begin{minipage}[c]{0.33\linewidth}
\centering
\includegraphics[width=1\textwidth]{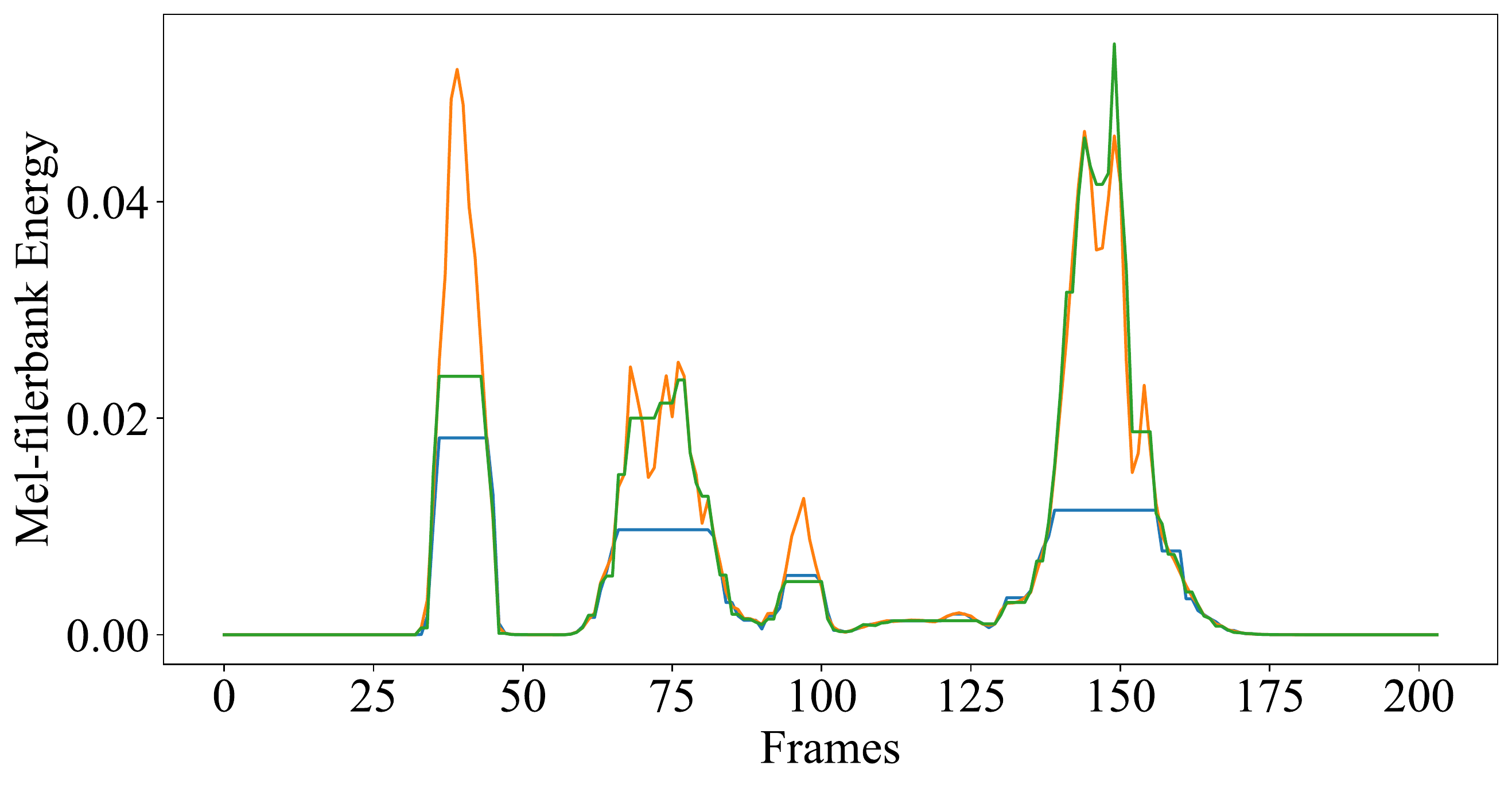}
\end{minipage}
}
\subfloat[Converted Sad]
{
\begin{minipage}[c]{0.33\linewidth}
\centering
\includegraphics[width=1\textwidth]{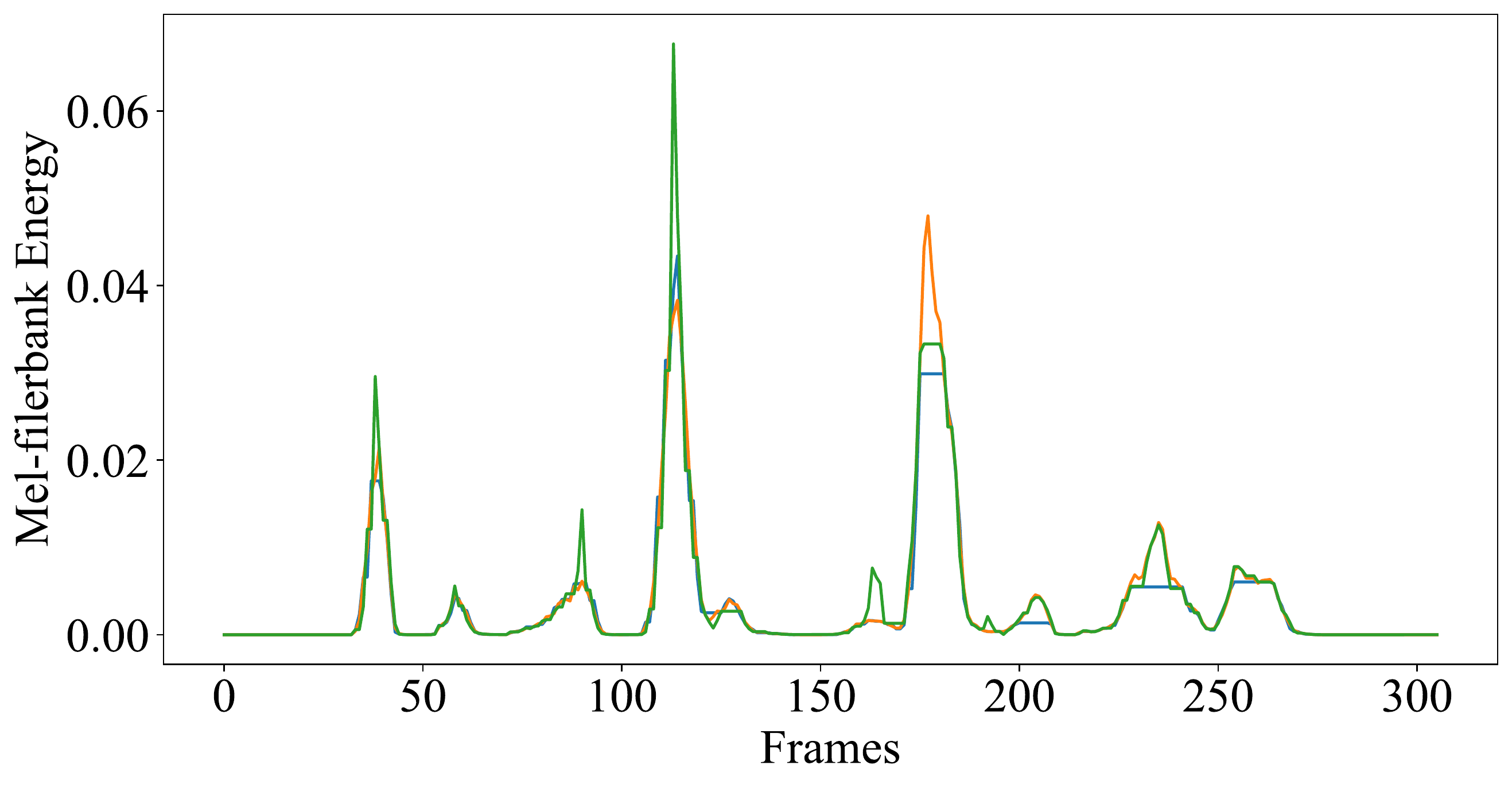}
\end{minipage}
}
\centering
\caption{A comparison of the speech energy from the emotional utterances converted by \textit{Emovox} with three different emotion intensities (0.1, 0.5 and 0.9).}
\label{fig:energy}
\end{figure*}
\subsection{Emovox versus Baselines}
\label{sec: baseline}

In this subsection, we include CycleGAN-EVC, StarGAN-EVC, and Seq2Seq-EVC as baselines. It is noted that these baselines do not have an intensity control module.  As a fair comparison, we conduct emotion intensity transfer for \textit{Emovox} in both objective and subjective evaluations. 

\subsubsection{Objective Evaluation}
{\noindent \bf Mel-cepstral Distortion (MCD):} 
As shown in Table \ref{tab:objective1}, all systems achieve better MCD values than that of the Zero Effort case. \textit{Zero Effort} case directly compares the source and target utterances without any conversion. We also observe that \textit{Emovox} completely outperforms CycleGAN-EVC and StarGAN-EVC. It also outperforms Seq2Seq-EVC for Neu-Ang and Neu-Hap 
(first three letters of source and target emotion, each) 
and achieves comparable results for Neu-Sad. This suggests that \textit{Emovox} is superior to the others in terms of spectrum conversion.

{\noindent \bf Differences of Duration (DDUR):} CycleGAN-EVC and StarGAN-EVC perform frame-by-frame mapping, but they do not convert the speech duration. Thus, the DDUR results of these two frameworks are not reported. 
As shown in Table \ref{tab:objective1}, compared with Seq2Seq-EVC, \textit{Emovox} achieves better results for both Neu-Ang and Neu-Hap for duration modelling, and achieves comparable results in Neu-Sad conversion. These results further confirm the effectiveness of \textit{Emovox} in terms of duration conversion.

\begin{table*}[t]
\centering
\caption{A comparison of best-worst scaling (BWS) test results for speech quality of three different emotion intensity control methods with \textit{Emovox}. }
\scalebox{0.90}{
\begin{tabular}{c|c|c|c|c|c|c|cc|c|c|c|c|c|c|c|c|c|c}
\hline
\multirow{3}{*}{Method } & \multicolumn{6}{c|}{Intensity = 0.1 (Weak)}                                                & \multicolumn{6}{c|}{Intensity = 0.5 (Medium)}                                                  & \multicolumn{6}{c}{Intensity = 0.9 (Strong)}                                              \\ \cline{2-19} 
                            & \multicolumn{2}{c|}{Neu-Ang} & \multicolumn{2}{c|}{Neu-Hap} & \multicolumn{2}{c|}{Neu-Sad} & \multicolumn{2}{c|}{Neu-Ang}      & \multicolumn{2}{c|}{Neu-Hap} & \multicolumn{2}{c|}{Neu-Sad} & \multicolumn{2}{c|}{Neu-Ang} & \multicolumn{2}{c|}{Neu-Hap} & \multicolumn{2}{c}{Neu-Sad} \\ \cline{2-19} 
                            & B             & W            & B             & W            & B             & W            & \multicolumn{1}{c|}{B}    & W    & B             & W            & B             & W            & B             & W            & B             & W            & B             & W            \\ \hline
Scaling                     & 8\%           & 31\%         & 15\%          & 15\%         & 8\%           & 70\%         & \multicolumn{1}{c|}{16\%} & 38\% & 8\%           & 8\%          & 8\%           & 62\%         & 17\%          & 31\%         & 9\%           & 6\%          & 8\%           & 69\%         \\ 
Attention                   & 15\%          & 69\%         & 0\%           & 77\%         & \textbf{54\%}          & \textbf{15\%}         & \multicolumn{1}{c|}{15\%} & 62\% & 0\%           & 92\%         & 43\%          & 31\%         & 6\%           & 69\%         & 0\%           & 92\%         & \textbf{54\%}          & 31\%         \\ 
\textbf{Relative}                    & \textbf{77\%}          & \textbf{0\%}          & \textbf{85\%}          & \textbf{8\%}          & 38\%          & \textbf{15\%}         & \multicolumn{1}{c|}{\textbf{69\%}} & \textbf{0\%}  & \textbf{92\%}          & \textbf{0\%}          & \textbf{49\%}          & \textbf{7\%}          & \textbf{77\%}          & \textbf{0\%}          & \textbf{91\%}          & \textbf{2\%}          & 38\%          & \textbf{0\%}          \\ \hline
\end{tabular}}
\begin{tablenotes}

\footnotesize
\item Note: Emovox w/ scaling factor, Emovox w/ attention weights, and Emovox w/ relative attributes are denoted as Scaling, Attention, and Relative respectively. ''B'' denotes ''Best'', and ''W'' denotes ''Worst''.
\end{tablenotes}
\label{tab:bws1}
\end{table*}

\begin{figure*}[t]
\captionsetup[subfigure]{justification=centering}
\centering
\subfloat[Emovox w/ Scaling Factor]
{
\begin{minipage}[c]{0.3\linewidth}
\centering
\includegraphics[width=0.85\textwidth]{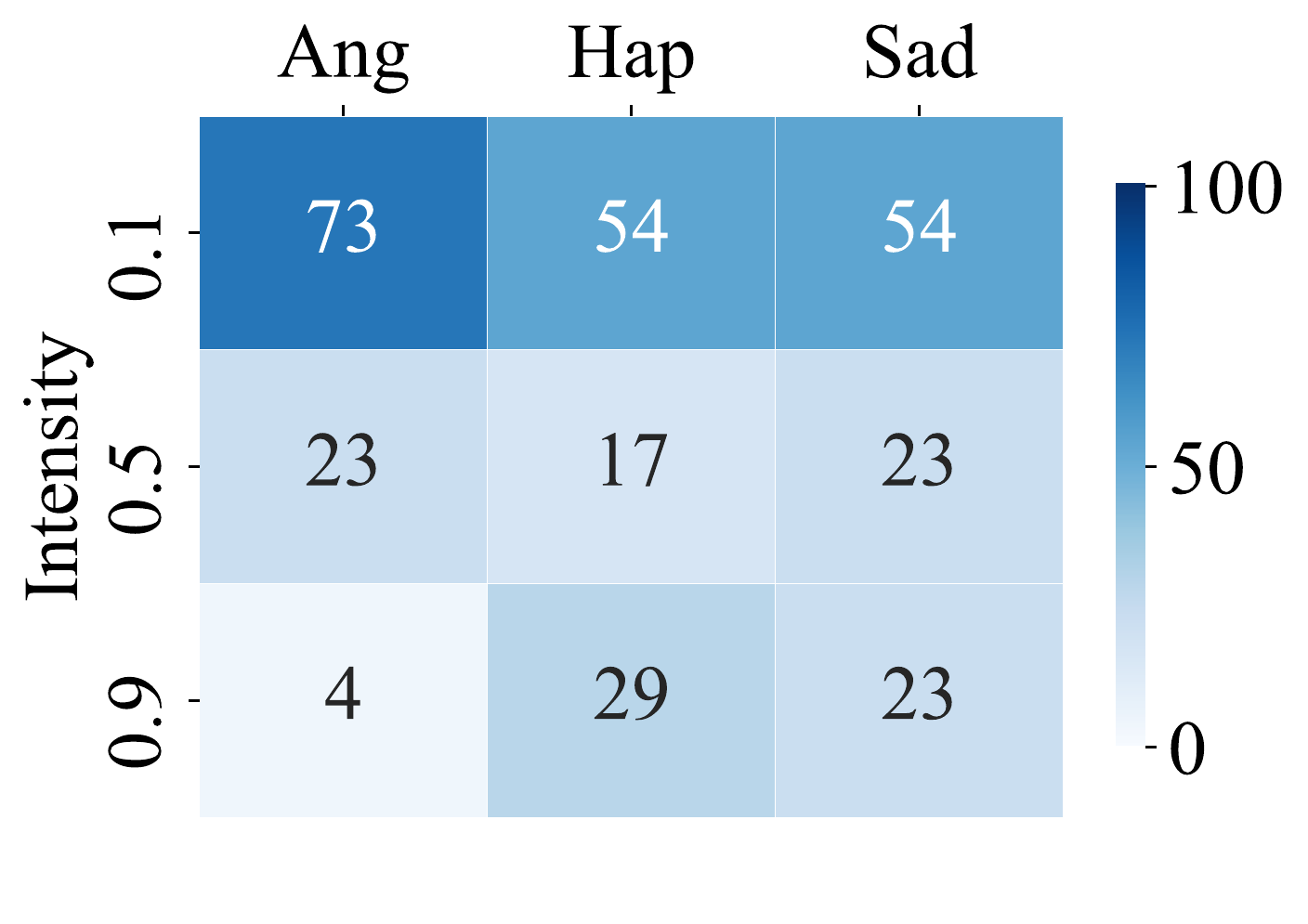}
\end{minipage}%
}
\subfloat[Emovox w/ Attention Weights]
{
\begin{minipage}[c]{0.3\linewidth}
\centering
\includegraphics[width=0.85\textwidth]{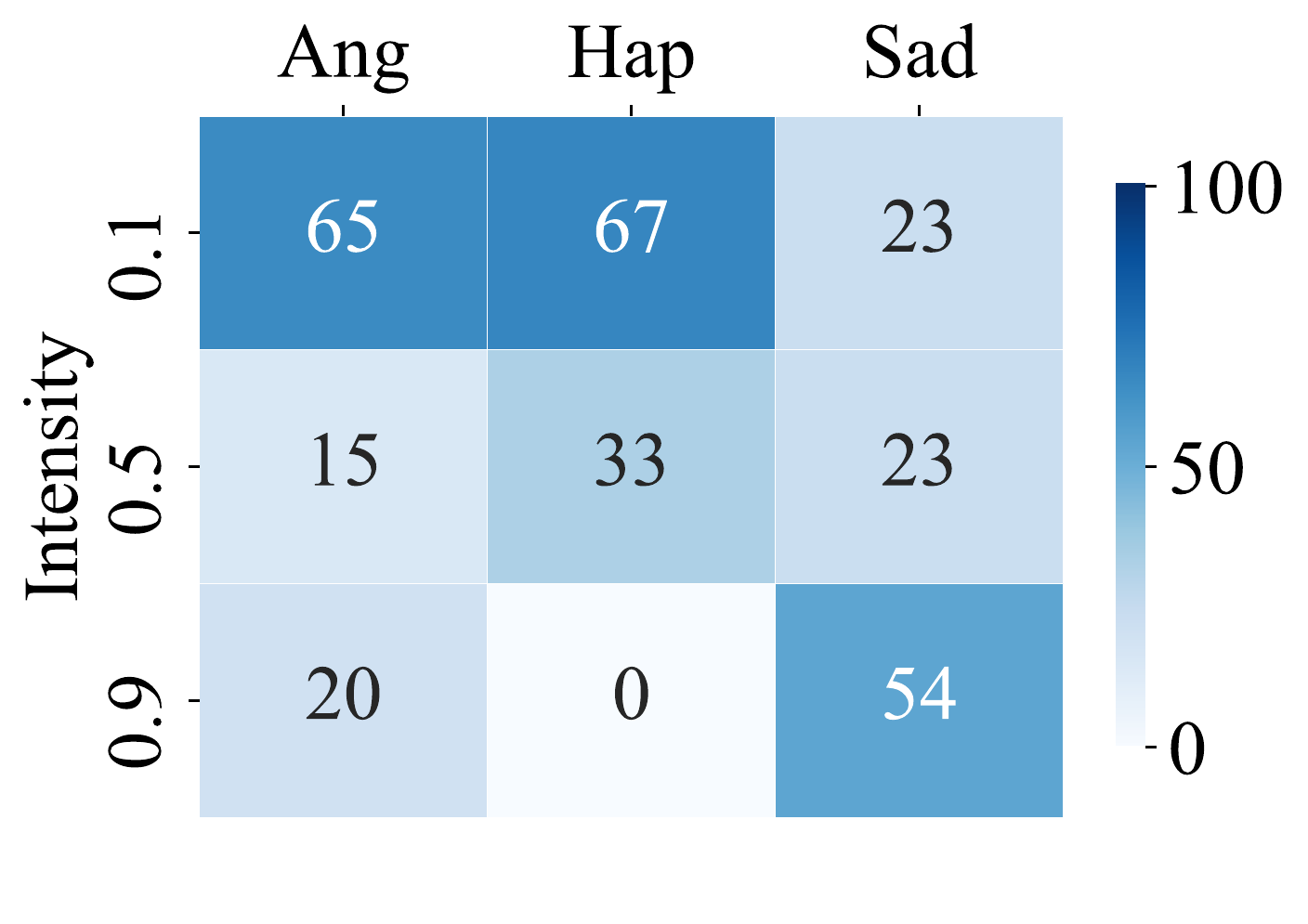}
\end{minipage}
}
\subfloat[Emovox w/ Relative Attributes]
{
\begin{minipage}[c]{0.3\linewidth}
\centering
\includegraphics[width=0.85\textwidth]{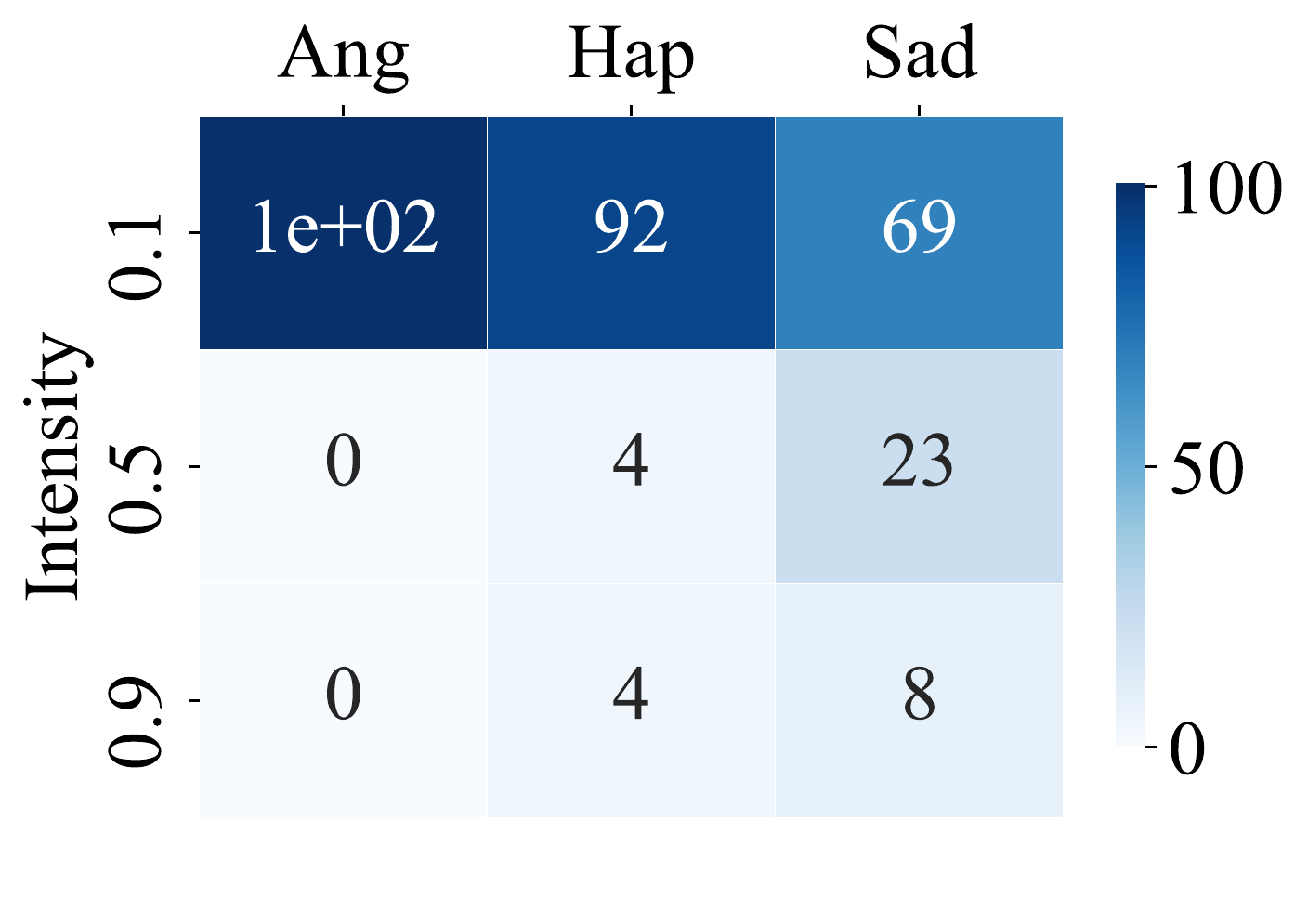}
\end{minipage}
}
\centering
\caption{A comparison of the preference percentage scores (\%) of Emovox with 3 different emotion intensity control methods. Listeners are asked to listen to the converted speech samples of 3 different emotion intensities (0.1, 0.5, and 0.9), and choose the least expressive one.}
\label{fig:leastbws}
\end{figure*}
\begin{figure*}[t]
\captionsetup[subfigure]{justification=centering}
\centering
\subfloat[Emovox w/ Scaling Factor]
{
\begin{minipage}[c]{0.3\linewidth}
\centering
\includegraphics[width=0.85\textwidth]{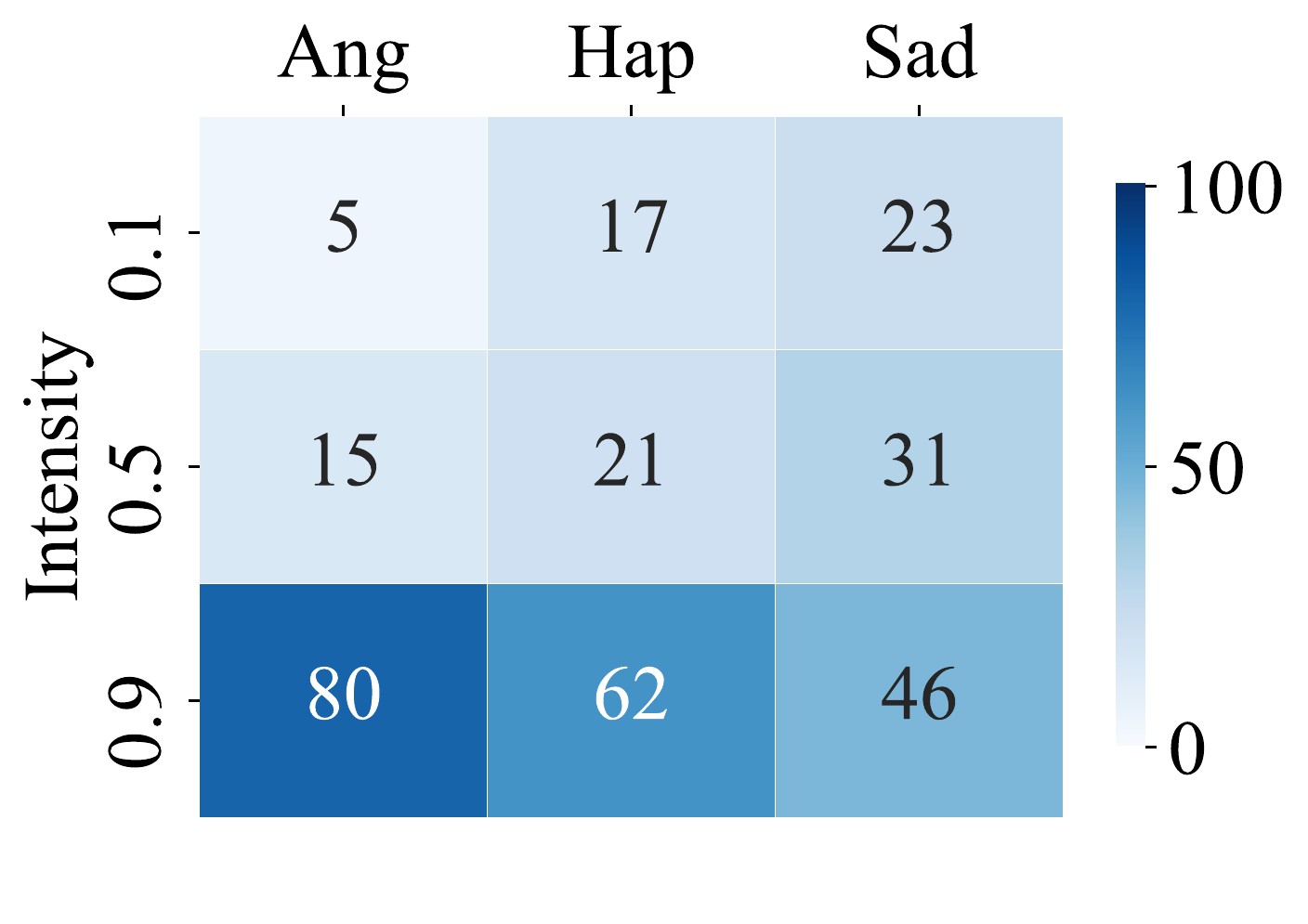}
\end{minipage}%
}
\subfloat[Emovox w/ Attention Weights]
{
\begin{minipage}[c]{0.3\linewidth}
\centering
\includegraphics[width=0.85\textwidth]{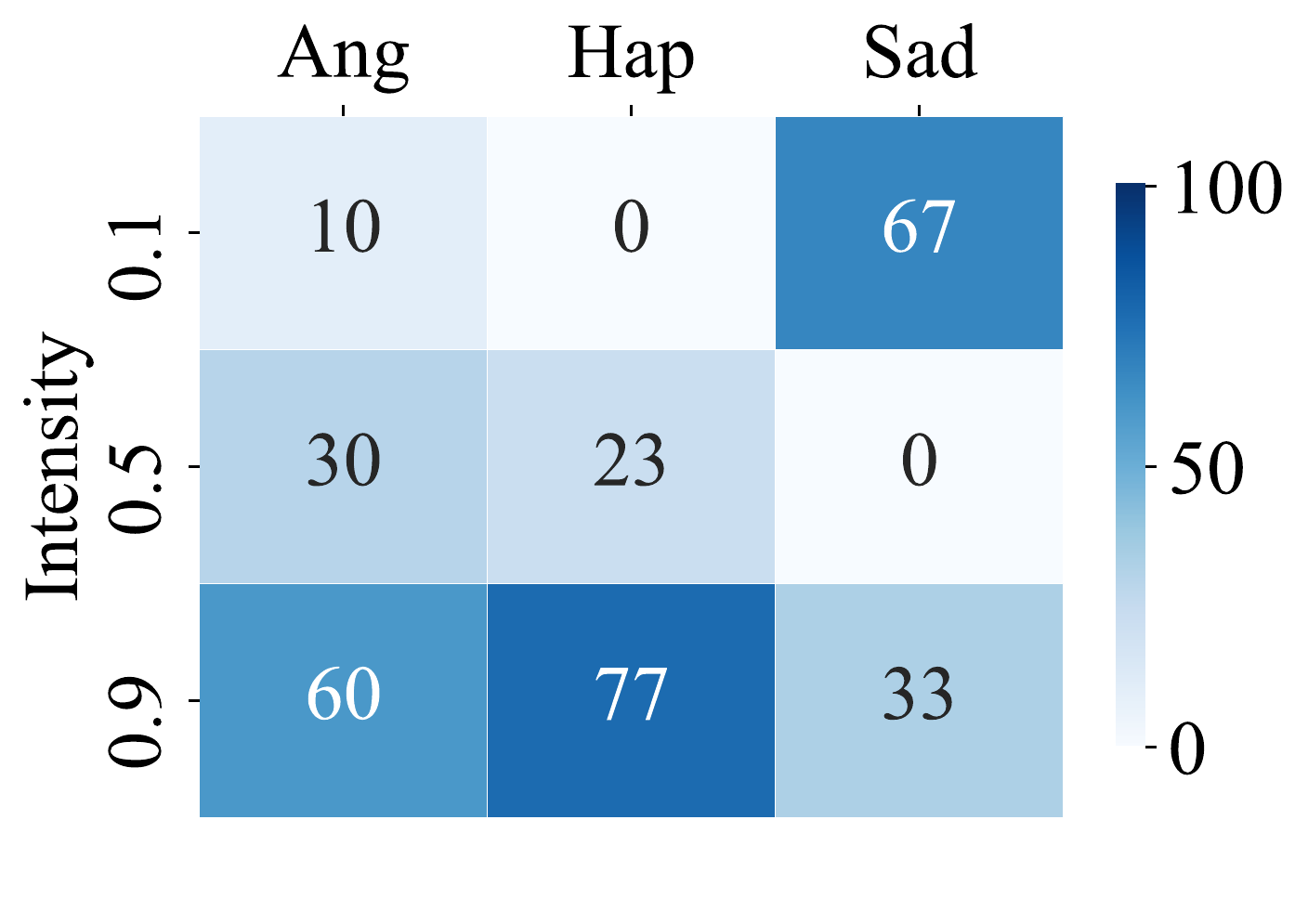}
\end{minipage}
}
\subfloat[Emovox w/ Relative Attributes]
{
\begin{minipage}[c]{0.3\linewidth}
\centering
\includegraphics[width=0.85\textwidth]{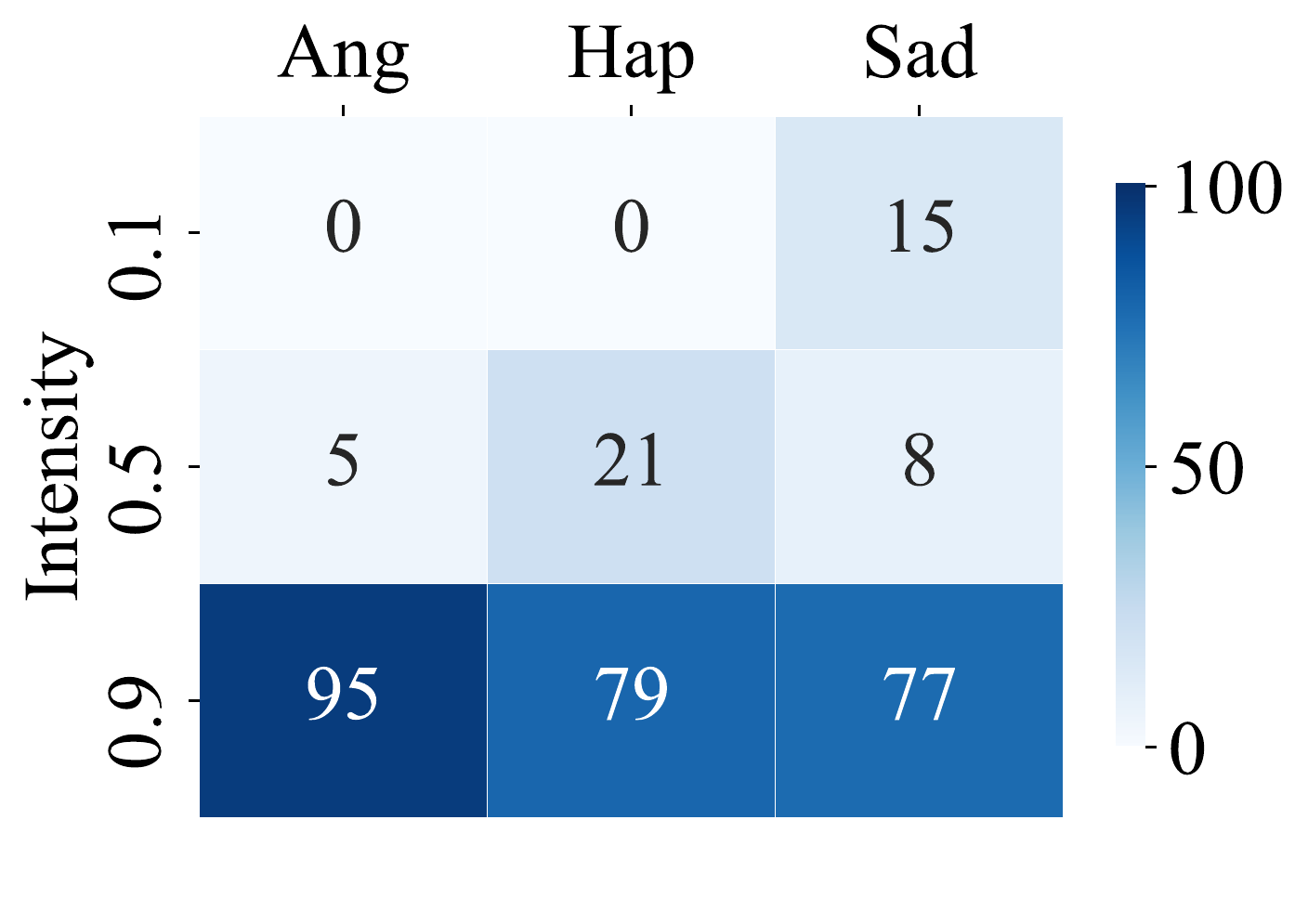}
\end{minipage}
}
\centering
\caption{A comparison of the preference percentage scores (\%) of \textit{Emovox} with 3 different emotion intensity control methods, and for 3 converted emotions. The subjects are asked to listen to the converted speech samples of 3 different emotion intensities (0.1, 0.5, and 0.9), and choose the most expressive one.}
\label{fig:mostbws}
\end{figure*}

\subsubsection{Subjective Evaluation}
\label{5.1.2}

We report Mean Opinion Score (MOS) test results for emotion similarity with the reference for our proposed \textit{Emovox} and all the baselines. From Figure 6, we observe that our proposed \textit{Emovox} consistently outperforms the baselines for all the emotion pairs. 
This observation is consistent with that in the objective evaluation. 
As for statistical significance, \textit{Emovox} achieves the most narrow confidence interval for Neu-Ang and Neu-Hap, that suggests a high level of consistency \cite{hosmer1992confidence}. Furthermore, we report the p-value of t-test scores of MOS between \textit{Emovox} and the others. We observe that almost all pairs achieve a p-value below $0.05$, confirming the significant results \cite{thisted1998p}. For Neu-Ang, the p-value between \textit{Emovox} and Seq2Seq-EVC is about $0.0558$, which is less than $0.1$ and still supports our claim.


\subsection{Emotion Intensity Control}
\label{sec: intensity control}

To evaluate the emotion intensity control in \textit{Emovox}, we choose three different intensity values: $0.1$, $0.5$, and $0.9$, corresponding to weak, medium, and strong. To understand the interplay between emotion intensity and different prosodic attributes, we first visualize several related prosodic cues of the converted emotional utterances with the same speaking content but different emotion intensities. We then compare our intensity control methods with other state-of-the-art methods. 

\subsubsection{Visual Comparisons}
\label{sec: visualization}
We visualize the prosodic attributes related to the emotion intensity, such as speech duration, pitch, and energy to gain an intuitive understanding of emotion intensity in vocal speech. Besides, we also would like to show that the emotion intensity control can be manifested in the changes of these prosodic features in our proposed framework. 

\begin{figure*}[t]
\centering
\subfloat[MCD results (a smaller value of MCD indicates a smaller spectral distortion)]
{
\begin{minipage}[c]{1\linewidth}
\centering
\includegraphics[width=11cm]{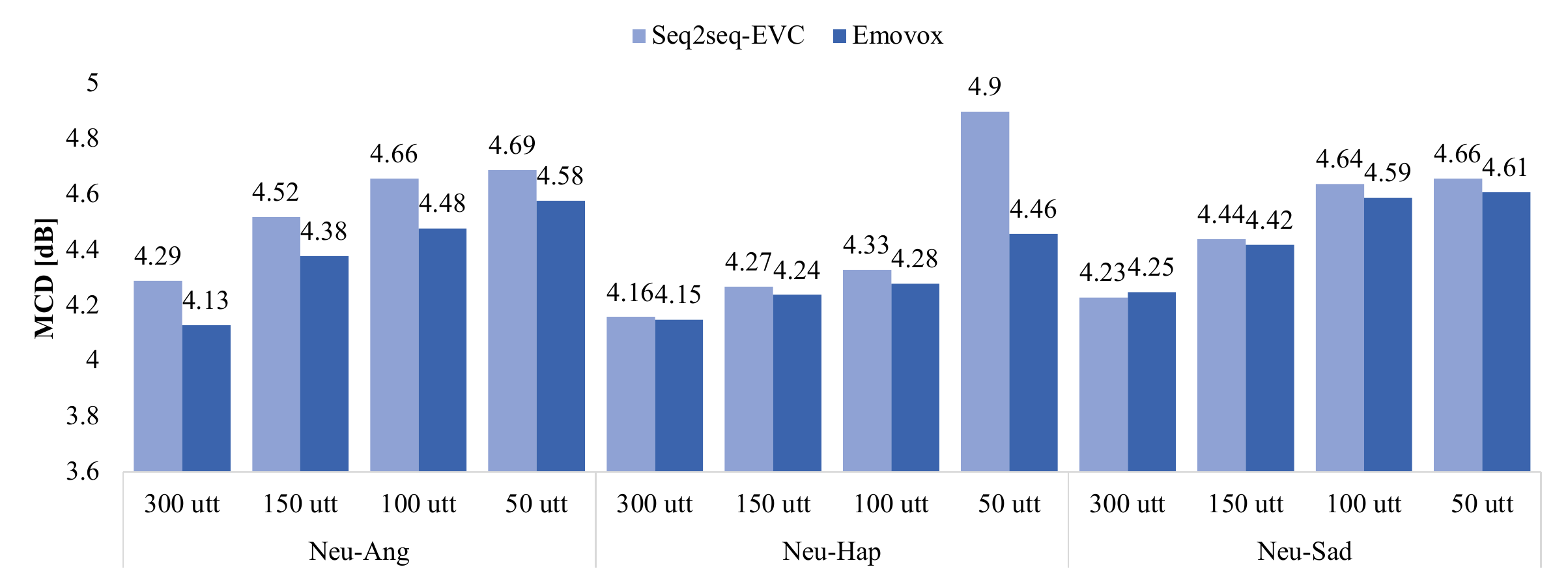}
\end{minipage}%
}
\newline

\subfloat[DDUR results (a smaller value of DDUR represents a better duration conversion performance) ]
{
\begin{minipage}[c]{1\linewidth}
\centering
\includegraphics[width=11cm]{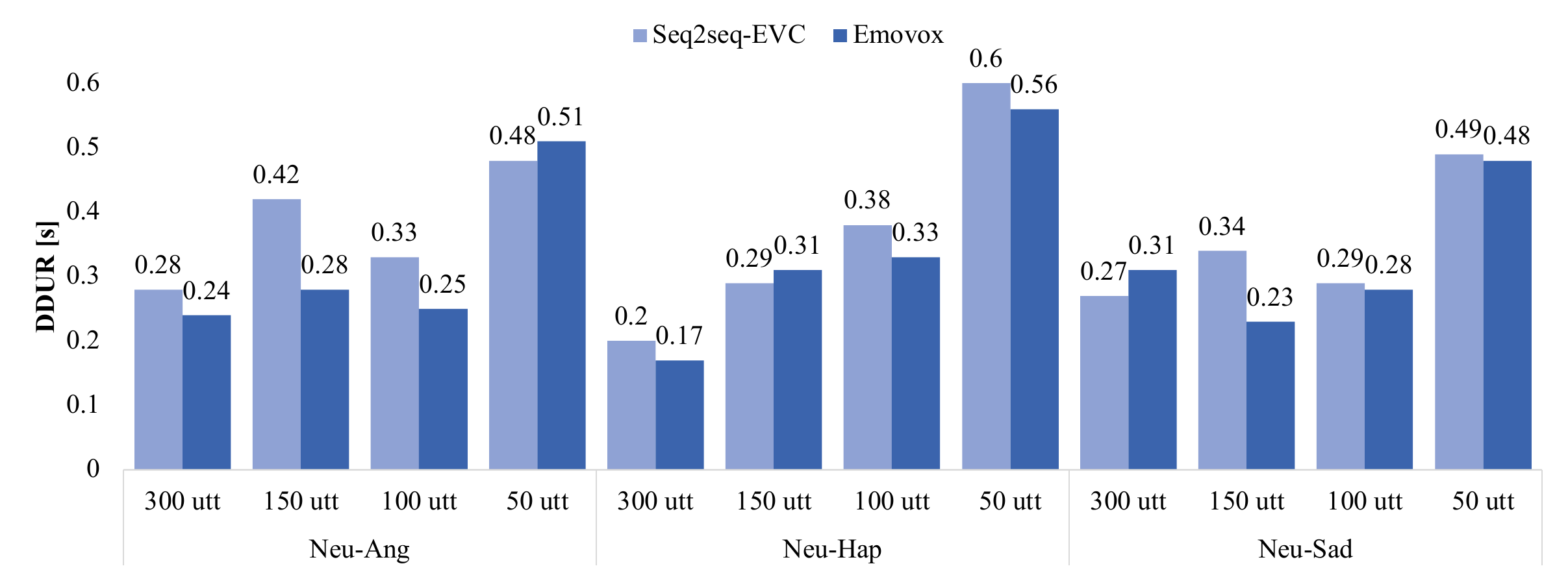}
\end{minipage}
}
\centering
\caption{A comparison of MCD and DDUR results of our proposed \textit{Emovox} and Seq2Seq-EVC for all three emotion pairs with different amounts of training utterances, that are 300, 150, 100, and 50 utterances, respectively. }
\label{fig:size}
\end{figure*}

\textbf{(1) Duration: }
Speech duration is considered as a distinct factor between active and passive emotions \cite{davitz1964personality}. To show that the emotion intensity is related to speaking rate, we compare the Mel-spectrogram of \textit{Sad} as a reference emotion, which is characterized with a slower speaking rate and more resonant timbre~\cite{owren2007measuring}, with that of its \textit{Neutral} emotion counterpart in Figure \ref{fig:ddur}(a). We also illustrate the Mel-spectrograms of  \textit{Emovox}-converted utterances with different intensities in Figure \ref{fig:ddur}(c),(d) and (e). 
We observe that the converted \textit{Sad} utterance with the highest intensity value has the slowest speaking rate among all three intensities (as shown in Figure \ref{fig:ddur}(e)). As the intensity value increases, the speaking rate decreases. 

\textbf{(2) Pitch Envelope: } Pitch envelope (i.\,e., the level, range, and shape of the pitch contour) is considered a major factor that contributes to the speech emotion, which is closely correlated to the activity level \cite{johnson1986recognition,owren2007measuring}. 
We represent pitch information with F0 contour, which is estimated with the harvest algorithm \cite{morise2017harvest} and aligned with dynamic time warping \cite{muller2007dynamic}.
In Figure \ref{fig:pitch}, we visualize the pitch contour of converted \textit{Angry}, \textit{Happy}, and \textit{Sad} utterances with three different intensities.
From Figure \ref{fig:pitch}(a) and (b), we observe that the converted \textit{Angry} and \textit{Happy} utterances with higher intensity values tend to have higher F0 values with larger fluctuations over time. This coincides the fact that the utterances with higher intensity values are more vibrant and sharper in expressing emotions such as \textit{Angry} and \textit{Happy}. For \textit{Sad}, there is no big difference in F0 range for different intensities, as shown in Figure \ref{fig:pitch}(c). This observation intuitively suggests that the intensity of expressing \textit{Sad} emotion may be more related to the speaking rate than the vocal pitch. 

\textbf{(3) Speech Energy: }
Speech energy measures the volume or the loudness of a voice \cite{scherer2003vocal,ververidis2006emotional}. Speech energy is often regarded as a prominent character of emotion intensity in the literature \cite{juslin2001impact,sobin1999emotion}. 
To show the effect of intensity control, we visualize and compare the energy contour of different intensities in Figure \ref{fig:energy}.
To represent the speech energy, we use 26 Mel-filterbanks and multiply each of them with the power spectrum. Then, we can measure the speech energy by adding up the coefficients.
As shown in Figure \ref{fig:energy}(a) and (b), we observe that the converted \textit{Angry} and \textit{Happy} utterances with higher intensity have lager energy values, which is consistent with our observations on the F0 contour. As for \textit{Sad}, we similarly observe that a higher intensity results in slightly higher energy values as shown in Figure \ref{fig:energy}(c). These observations show that our proposed \textit{Emovox} can effectively control the emotion intensity manifested in multiple prosodic factors in vocal speech.

\begin{table*}[t]
\centering
\caption{The effect of the perceptual loss function for the emotion similarity in a  best-worst scaling (BWS) test for four variants of Emovox w/o Intensity framework.} 
\scalebox{1}{
\begin{tabular}{c|c|c|c|c|c|c}
\hline
\multirow{2}{*}{Emovox w/o Intensity} & \multicolumn{2}{c|}{Neu-Ang} & \multicolumn{2}{c|}{Neu-Hap} & \multicolumn{2}{c}{Neu-Sad} \\ \cline{2-7} 
                           & Best          & Worst        & Best          & Worst        & Best          & Worst        \\ \hline
w/ $L_{sim}$ and $L_{cls}$                  &  \textbf{56\%}     &   \textbf{5\%}     &   \textbf{49\%}      &   \textbf{13\%}     &   \textbf{33\%}     &  16\%     \\ 
w/ $L_{sim}$            &   33\%     &  22\%     &  47\%       &  13\%       &   25\%       &  \textbf{11\%}    \\ 
w/ $L_{cls}$            &  9\%   &  24\%       &   2\%      &  24\%     &    22\%     &   22\%  \\ 
w/o $L_{sim}$ or $L_{cls}$      &  2\%       &  49\%    &  2\%       &  49\%     &   20\%       &  51\%     \\ \hline
\end{tabular}}
\label{tab:ablation}
\end{table*}

\subsubsection{Comparison with State-of-the-Art Control Methods}
\label{sec: compare}
As a comparative study, we implement three intensity control methods (i.\,e., Emovox w/ scaling factor, Emovox w/ attention weights, and Emovox w/ relative attributes) as described in Section \ref{sec: reference}. We evaluate the performance of these three methods in terms of speech quality and intensity control.

\textbf{(1) Speech Quality:} We first report the BWS listening test on \textit{Emovox} for speech quality evaluation in Table \ref{tab:bws1}.
At each intensity value, the subjects are asked to evaluate the speech quality of the converted emotional speech with 3 different emotion intensity control methods.

From Table \ref{tab:bws1}, we observe that \textit{Emovox} w/ relative attributes always achieves the best results for Neu-Ang and Neu-Hap, and comparable results with \textit{Emovox} w/ attention weights for Neu-Sad. These results show that \textit{Emovox} w/ relative attributes can achieve better speech quality while controlling the output emotion intensity than other control methods.

\textbf{(2) Intensity Control:} We then report another BWS test to evaluate the performance of emotion intensity control. 
For each framework, listeners are asked to assess the emotional expressiveness among three different intensities. 
We conjecture that the speech samples with an intensity value of $0.9$ sound more expressive than others, while those with an intensity value of $0.1$ sound more neutral. We report the preference percentage scores (\%) of the most and the least expressiveness for each controlling method in Figure \ref{fig:mostbws} and Figure \ref{fig:leastbws}, respectively.

As illustrated in Figure \ref{fig:mostbws}(c), \textit{Emovox} w/ relative attributes achieve the best preference results on intensity control, where most listeners choose the samples with an intensity value of $0.9$ as the most expressive ones. We also note that \textit{Emovox} w/ scaling factor and \textit{Emovox} w/ attention weights work well for converted angry and happy, as shown in Figure \ref{fig:mostbws}(a) and (b). However, their performance of converted sad is not satisfactory. We further observe that \textit{Emovox} w/ relative attributes also work better than the others, where most listeners choose the samples with an intensity value of $0.1$ as the least expressive ones, as shown in Figure \ref{fig:leastbws}(c). This observation is consistent with the previous one, which further validates the superior performance of relative attributes on emotion intensity control.

As a summary, our proposed \textit{Emovox} w/ relative attributes shows better performance on emotion intensity control while achieving better speech quality than other control methods.


\subsection{Impact of Training Data Size}
\label{sec: size}


To evaluate the effect of training data on the final performance, we gradually reduce the number of utterances used at the emotion training stage. We use $300$, $150$, $100$, and $50$ training utterances for each emotion, and use $20$ utterances for evaluation. In Figure \ref{fig:size}, we report the MCD and DDUR results of \textit{Emovox} and the baseline Seq2Seq-EVC. 

We observe that \textit{Emovox} consistently achieves better MCD results than Seq2Seq-EVC. We further observe that the MCD scores for \textit{Emovox} between $150$ to $50$ training utterances are comparable and not significantly poorer than that of using $300$ utterances. This indicates \textit{Emovox}'s robustness to limited training data.

For DDUR, we first observe that both \textit{Emovox} and Seq2Seq-EVC have much higher DDUR values with $50$ training utterances. It suggests that both frameworks cannot predict the speech duration well if the training size is too small. Between $300$ to $100$ training utterances, the performance of \textit{Emovox} is comparable, which again attest to \textit{Emovox}'s robustness to limited training data.

\subsection{Ablation Studies}
\label{sec: ablation}
We conduct ablation studies to validate the contributions of 1) style pre-training, and 2) perceptual losses from pre-trained SER in emotion training.

\subsubsection{Style Pre-training} We first compare the  Mel-cepstral  distortion (MCD) results of \textit{Emovox} and \textit{Emovox} (w/o style pre-training), where the latter is trained directly with a limited amount of emotional speech data and without any pre-training. As shown in Table \ref{tab:objective1}, \textit{Emovox} (w/o style pre-training) provides the worst results for all emotion pairs. 

We further compare \textit{Emovox} and \textit{Emovox} (w/o style pre-training) in terms of DDUR as shown in Table \ref{tab:objective1}. We observe that \textit{Emovox} (w/o style pre-training) has the worst DDUR results. These results validate the effectiveness of style pre-training. 

\subsubsection{Perceptual Loss Functions}
\label{sec: loss}
As discussed in Section \ref{sec: perceptual},
we expect that the emotion embedding similarity loss $L_{sim}$ and emotion classification loss $L_{cls}$ help generate more discriminative embeddings (see Figure \ref{fig:embedding}). To further validate the effectiveness of these two loss functions on the final performance, we conduct a best-worse scaling listening test where we evaluate the emotion similarity with the reference emotion. 
To be consistent with Section \ref{sec: perceptual}, we only conduct an ablation study with the \textit{Emovox w/o intensity} configuration.
The results are reported in Table \ref{tab:ablation}.

From Table \ref{tab:ablation}, we observe that most listeners choose ``\textit{Emovox w/o intensity, w/ $L_{sim}$ and $L_{cls}$}'' as the best in terms of emotion similarity, while most of them choose ``\textit{Emovox w/o intensity, w/o $L_{sim}$ or $L_{cls}$}'' as the worst for all the emotion pairs. This suggests that these two loss functions improve emotional expressiveness, which validates the idea of incorporating SER losses for emotion supervision.

\section{Conclusion}
\label{sec: concludes}

This contribution filled the research gap of emotion intensity control in current emotional voice conversion literature.
We proposed a novel emotional voice conversion framework --  \textit{Emovox} -- that is based on a sequence-to-sequence model. The proposed \textit{Emovox} framework provides a fine-grained, effective emotion intensity control for the first time in emotional voice conversion. The key highlights are as follows:
\begin{enumerate}
   \item We formulated an emotion intensity modeling technique and proposed an emotion intensity controlling mechanism based on relative attributes. We proved that our proposed mechanism outperformed other competing controlling methods in speech quality and emotion intensity control.
    
    \item Instead of simply correlating emotion intensity with the loudness of a voice, we presented a comprehensive analysis for the first time to understand the interplay between emotion intensity and various prosodic attributes such as speech duration, pitch envelope, and speech energy. We showed that our emotion intensity control could be manifested in various prosodic aspects. 
     \item We proposed style pre-training and perceptual losses from a pre-trained SER to improve the emotion intelligibility in converted emotional speech. We showed that \textit{Emovox} outperformed state-of-the-arts emotional voice conversion frameworks. 
     With style pre-training and perceptual losses from a pre-trained SER, \textit{Emovox} could effectively perform well with a limited amount of emotional speech data.
 \end{enumerate}

Our future directions include the study of cross-lingual emotional voice conversion and emotion style modelling with self-supervised learning.
In addition, a closer coupling of conversion and speech emotion recognition is foreseen: conversion can help augment training data for recognition, while recognition can serve as objective conversion training guidance. 

\ifCLASSOPTIONcompsoc
  \section*{Acknowledgments}
\else
  \section*{Acknowledgment}
\fi

The research by Kun Zhou and Haizhou Li is supported by the Science and Engineering Research Council, Agency of Science, Technology and Research (A*STAR), Singapore, through the National Robotics Program under Human-Robot Interaction Phase 1 (Grant No.\ 192 25 00054);  Human Robot Collaborative AI under its AME Programmatic Funding Scheme (Project No. A18A2b0046); National Research Foundation Singapore under its AI Singapore Programme (Award Grant No: AISG-100E-2018-006), and by A*STAR under its RIE2020 Advanced Manufacturing and Engineering Domain (AME) Programmatic Grant (Grant No.\ A1687b0033, Project Title: Spiking Neural Networks).

The research by Berrak Sisman is supported by the Ministry of Education, Singapore, under its MOE Tier 2 funding programme, award no: MOE-T2EP50220-0021, SUTD Start-up Grant Artificial Intelligence for Human Voice Conversion (SRG ISTD 2020 158) and SUTD AI Grant -- Thrust 2 Discovery by AI (SGPAIRS1821).

The authors would like to thank the anonymous reviewers for their insightful comments, Dr Bin Wang for valuable discussions and Dr Rui Liu for sharing part of the codes.

\ifCLASSOPTIONcaptionsoff
  \newpage
\fi

\bibliographystyle{IEEEtran}
\bibliography{Bibliography}

\begin{IEEEbiography}[{\includegraphics[width=1in,height=1.25in,clip,keepaspectratio]{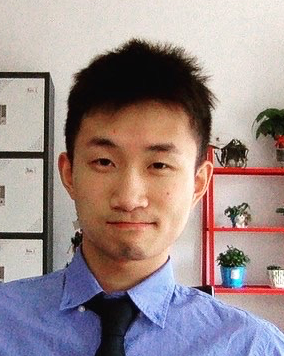}}]{Kun Zhou}(Student Member, IEEE)
received his B.\,Eng.\ degree from the School of Information and Communication Engineering, University of Electronic Science and Technology of China (UESTC), Chengdu, China, in 2018, and the M.\,Sc.\ degree from the Department of Electrical and Computer Engineering, National University of Singapore (NUS), Singapore, in 2019. He is currently a PhD student at the National University of Singapore. 
His research interests mainly focus on emotion analysis and synthesis in speech, including emotional voice conversion and emotional text-to-speech. He served as local arrangment co-chair of IEEE ASRU 2019, SIGDIAL 2021, IWSDS 2021, O-COCOSDA 2021 and ICASSP 2022. He is a reviewer of ICASSP and Speech Communication.
\end{IEEEbiography}
\vskip -2\baselineskip plus -1fil
\begin{IEEEbiography}[{\includegraphics[width=1in,height=1.25in,clip,keepaspectratio]{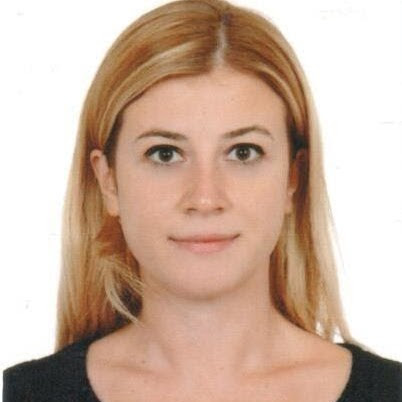}}]{Berrak Sisman}(Member, IEEE)
received her PhD degree in Electrical and Computer Engineering from National University of Singapore in 2020, fully funded by the A*STAR Graduate Academy under the Singapore International Graduate Award (SINGA). She is currently working as an Assistant Professor at the Singapore University of Technology and Design (SUTD). She is also an Affiliated Researcher at the National University of Singapore (NUS). Prior to joining SUTD, she was a Postdoctoral Research Fellow at the National University of Singapore, and a Visiting Researcher at Columbia University, New York, United States. She was also an exchange PhD student at the University of Edinburgh and a visiting scholar at The Centre for Speech Technology Research (CSTR), University of Edinburgh, in 2019. She was attached to the RIKEN Advanced Intelligence Project, Japan in 2018. Her research is focused on machine learning, signal processing, speech synthesis, voice conversion, and emotion. She has served as the General Coordinator of the Student Advisory Committee (SAC) of the International Speech Communication Association (ISCA), and is currently serving as the General Coordinator of the ISCA Postdoc Advisory Committee (PECRAC). She is appointed as an Area Chair (Speech Synthesis) at INTERSPEECH 2021 and 2022, and Publication Chair of ICASSP 2022. She is elected as a member of the IEEE Speech and Language Processing Technical Committee (SLTC) in the area of Speech Synthesis for the term from 2022 to 2024.
\end{IEEEbiography}
\vskip -2\baselineskip plus -1fil

\begin{IEEEbiography}[{\includegraphics[width=1in,height=1.25in,clip,keepaspectratio]{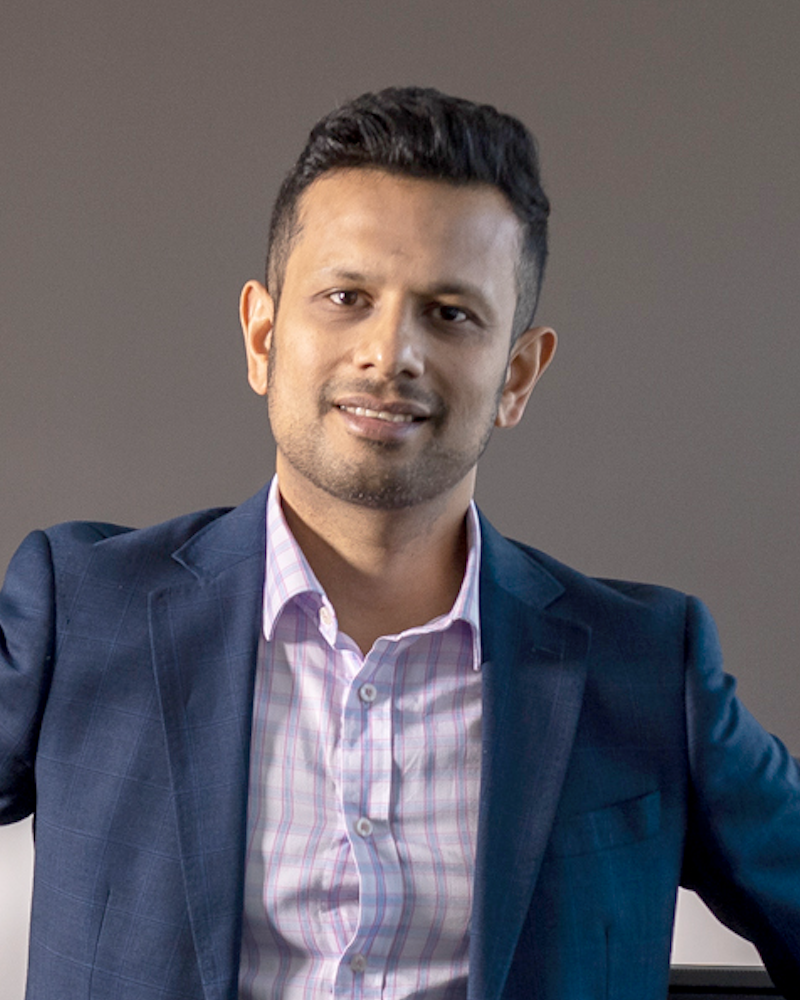}}]{Rajib Rana}(Member, IEEE)
received the B.\,Sc.\ degree in computer science and engineering from Khulna University, with the Prime Minister and President's Gold Medal for outstanding achievements, and the Ph.\,D.\ degree in computer science and engineering from the University of New South Wales, Sydney, Australia, in 2011. He received his Postdoctoral Training with the Autonomous System Laboratory, CSIRO, before joining the University of Southern Queensland, as a Faculty Member, in 2015. 
He is currently a Senior Advance Queensland Research Fellow and an Associate Professor with the University of Southern Queensland. He is also the Director of the 
IoT Health Research Program with the University of Southern Queensland, which capitalises on advancements in technology and sophisticated information and data processing to understand disease progression in chronic health conditions better and develop predictive algorithms for chronic diseases, such as mental illness and cancer. His current research interests include unsupervised representation learning, Adversarial Machine Learning, Re-enforcement Learning, Federated Learning, Emotional Speech Generation, and Domain Adaptation.
\end{IEEEbiography}
\vskip -2\baselineskip plus -1fil
\begin{IEEEbiography}[{\includegraphics[width=1in,height=1.25in,clip,keepaspectratio]{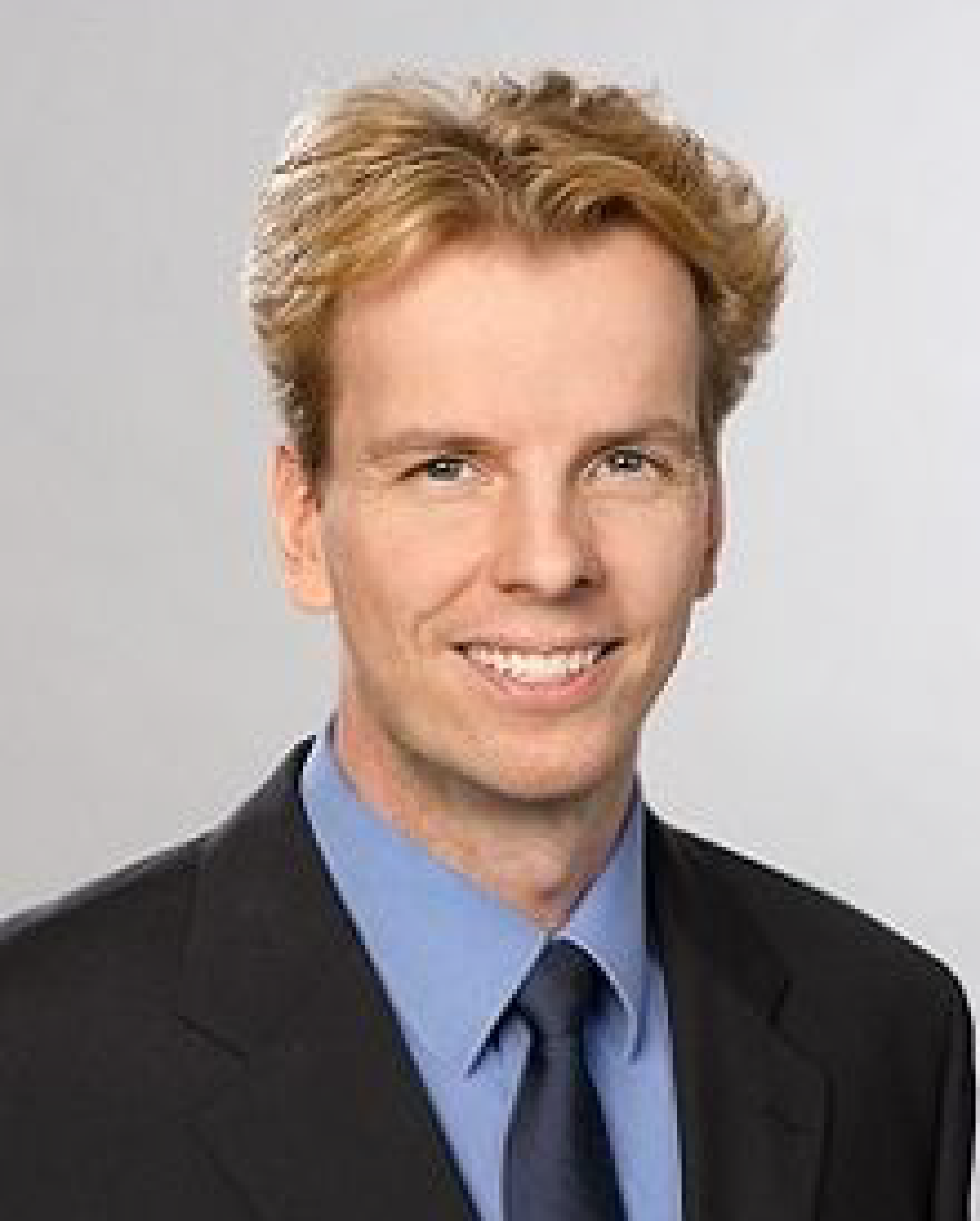}}]{Bj{\"o}rn W. Schuller}(M’05-SM’15-F’18)
received the diploma degree, the doctoral degree in automatic speech and emotion recognition, and the habilitation and Adjunct Teaching Professor in signal processing and machine intelligence from Technische Universit{\"a}t M{\"u}nchen (TUM), Munich, Germany, in 1999, 2006, and 2012, respectively, all in electrical engineering and information technology. He is currently a Professor of Artificial Intelligence with the Department of Computing, Imperial College London, U.K., where he heads the Group on Language, Audio, and Music (GLAM), a Full Professor and the Head of the Chair of Embedded Intelligence for Health Care and Wellbeing with the University of Augsburg, Germany, and the founding CEO/CSO of audEERING. He was previously a Full Professor and the Head of the Chair of Complex and Intelligent Systems with the University of Passau, Germany. He has (co-)authored five books and more than 1\,000 publications in peer-reviewed books, journals, and conference proceedings leading to more than overall 40,000 citations (H-index=96). He was an Elected Member of the IEEE Speech and Language Processing Technical Committee. He is a Golden Core Member of the IEEE Computer Society, a Fellow of the IEEE, AAAC, BCS, and ISCA, as well as a Senior Member of the ACM, and the President-Emeritus of the Association of the Advancement of Affective Computing (AAAC). He was the General Chair of ACII 2019, a Co-Program Chair of Interspeech, in 2019, and ICMI, in 2019, a repeated Area Chair of ICASSP, next to a multitude of further Associate and a Guest Editor roles and functions in Technical and Organisational Committees. He is the Field Chief Editor of the Frontiers in Digital Health and a former Editor-in-Chief of the IEEE Transcations on Affective Computing. 
\end{IEEEbiography}
\vfill
\begin{IEEEbiography}[{\includegraphics[width=1in,height=1.25in,clip,keepaspectratio]{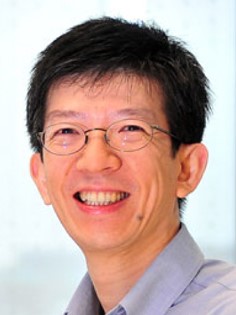}}]{Haizhou Li}
(M’91-SM’01-F’14) received the B.\,Sc., M.\,Sc., and Ph.D degree in electrical and electronic engineering from South China University of Technology, Guangzhou, China in 1984, 1987, and 1990 respectively. Dr Li is currently a Professor at the School of Data Science, The Chinese University of Hong Kong (Shenzhen), China, and the Department of Electrical and Computer Engineering, National University of Singapore (NUS). His research interests include automatic speech recognition, speaker and language recognition, and natural language processing. Prior to joining NUS, he taught in the University of Hong Kong (1988-1990) and South China University of Technology (1990-1994). He was a Visiting Professor at CRIN in France (1994-1995), Research Manager at the Apple-ISS Research Centre (1996-1998), Research Director in Lernout \& Hauspie Asia Pacific (1999-2001), Vice President in InfoTalk Corp.\ Ltd.\ (2001-2003), and the Principal Scientist and Department Head of Human Language Technology in the Institute for Infocomm Research, Singapore (2003-2016). Dr Li served as the Editor-in-Chief of IEEE/ACM Transactions on Audio, Speech and Language Processing (2015-2018), and as a Member of the Editorial Board of Computer Speech and Language (2012-2018). He was an elected Member of IEEE Speech and Language Processing Technical Committee (2013-2015), the President of the International Speech Communication Association (2015-2017), the President of the Asia Pacific Signal and Information Processing Association (2015-2016), and the President of the Asian Federation of Natural Language Processing (2017-2018). He was the General Chair of ACL 2012, INTERSPEECH 2014, and ASRU 2019. Dr Li is a Fellow of the IEEE and the ISCA. He was a recipient of the National Infocomm Award 2002 and the President's Technology Award 2013 in Singapore. He was named one of the two Nokia Visiting Professors in 2009 by the Nokia Foundation, and Bremen Excellence Chair Professor in 2019.

\end{IEEEbiography}




\end{document}